\documentclass[a4paper]{article}

\usepackage{xcolor}
\usepackage{stmaryrd}
\usepackage{latexsym}
\usepackage{amsmath}
\usepackage{amsfonts}
\usepackage{amssymb}
\usepackage{bm}
\usepackage{graphicx}
\usepackage{hyperref}
\usepackage{cite}
\usepackage{tikz}
\usepackage{pgfplots, booktabs}
\usepgfplotslibrary{groupplots}

\DeclareMathAlphabet{\mathpzc}{OT1}{pzc}{m}{it}

\newcommand{\be}{\begin{equation}}
\newcommand{\ee}{\end{equation}}

\newcommand{\ef}[1]{\, #1}

\newcommand{\vx}{\vec{x}}
\newcommand{\vy}{\vec{y}}
\newcommand{\vs}{\vec{s}}
\newcommand{\sig}{\sigma}

\newcommand{\bR}{\mathbb{R}}

\newcommand{\bT}{\mathbb{T}}

\newcommand{\dx}[1] {\!\mathrm{d}{#1}\,}
\newcommand{\ddx}[1] {\,\mathrm{d}{#1}}

\newcommand{\cO}{\mathcal{O}}

\newcommand{\eee}{\mathbb{E}}

\newcommand{\eps}{\epsilon}
\newcommand{\lam}{\lambda}

\newcommand{\smfrac}[2]{\genfrac{}{}{0.25pt}{1}{#1}{#2}}

\newcommand{\pio}{\pi_{\rm opt}}
\newcommand{\pid}{\pi_{\rm Dyck}}
\newcommand{\Ho}{H_{\rm opt}}
\newcommand{\Hd}{H_{\rm Dyck}}

\newcommand{\deenne}[2]{\frac{\partial^#2}{\partial #1 ^#2}}

\newcommand{\kS}{\mathfrak{S}}

\newcommand{\kb}{k_{\rm blue}}
\newcommand{\kr}{k_{\rm red}}
\newcommand{\hb}{h_{\rm blue}}
\newcommand{\hr}{h_{\rm red}}

\newtheorem{theorem}{Theorem}
\newtheorem{remark}{Remark}
\newtheorem{corollary}{Corollary}
\newtheorem{conj}{Conjecture}
\newtheorem{lemma}{Lemma}

\newtheorem{proposition}{Proposition}
\newtheorem{definition}{Definition}

\newcommand{\pf}{\noindent {\it Proof.}}
\newcommand{\qed}{\hfill $\square$}

\begin{document}

\title{The Dyck bound in the concave 1-dimensional random assignment model
  }

  \author{Sergio Caracciolo$^{1}$, Matteo P. D'Achille$^{2}$, Vittorio Erba$^{1}$\\  and Andrea
  Sportiello$^{3}$
\\[0.5cm]
\normalsize 
\rule{0pt}{16pt}%
$^{1}$~Dipartimento di Fisica dell'Universit\`a di Milano, and
  INFN, sez.~di Milano,\\
\normalsize 
  Via Celoria 16, 20100 Milano, Italy.\\
\normalsize 
$^{2}$Centre CEA de Saclay, Gif-sur-Yvette, France, CIRB Coll\`ege de France, \\ \normalsize
11 Place Marcelin Berthelot, 75231 Paris,  \\ \normalsize
 LI-PaRAD Universit\'e de Versailles Saint-Quentin-en-Yvelines, Versailles \\ \normalsize
and Universit\'e Paris Saclay, France\\ 
\normalsize
$^{3}$~LIPN, and CNRS, Universit\'e Paris 13, Sorbonne Paris Cit\'e,\\
\normalsize 
  99 Av.~J.-B.~Cl\'ement, 93430 Villetaneuse, France.
\\
\\
\normalsize doi:10.1088/1751-8121/ab4a34
}

\date{\today}

\maketitle

\begin{abstract} 
\noindent
We consider models of assignment for  random $N$ blue points and $N$
red points on an interval of length $2N$, in which the cost for
connecting a blue point in $x$ to a red point in $y$ is the concave
function $|x-y|^p$, for $0<p<1$.  Contrarily to the convex case $p>1$,
where the optimal matching is trivially determined, here the
optimization is non-trivial.

The purpose of this paper is to introduce a special configuration,
that we call the \emph{Dyck matching}, and to study its statistical
properties. We compute exactly the average cost, in the asymptotic
limit of large $N$, together with the first subleading correction. The
scaling is remarkable: it is of order $N$ for $p<\frac{1}{2}$, order
$N \ln N$ for $p=\frac{1}{2}$, and $N^{\frac{1}{2}+p}$ for
$p>\frac{1}{2}$, and it is universal for a wide class of models.  
We conjecture that the average cost of the Dyck
matching has the same scaling in $N$ as the cost of the optimal
matching, and we produce numerical data in support of this
conjecture. We hope to produce a proof of this claim in future work.
\end{abstract}

\vfil\eject

\section{The problem}

\subsection{Models of Random Assignment}

In this paper we study the statistical properties of the
\emph{Euclidean random assignment problem}, in the case in which the
points are confined to a one-dimensional interval, and the cost is a
\emph{concave} increasing function of their distance.

The \emph{assignment problem} is a combinatorial optimization problem,
a special case of the \emph{matching problem} when the underlying
graph is bipartite.  As for any combinatorial optimization problem,
each realisation of the problem is described by an \emph{instance}
$J$, and the goal is to find, within some space of configurations, the
particular one that minimises the given cost function.  In the
assignment problem, the instance $J$ is a real-positive $N \times N$
matrix, encoding the costs of each possible pairing among $N$ blue and
$N$ red points ($J_{ij}$ is the cost of pairing the $i$-th blue point
with the $j$-th red point), the space of configurations is the set of
permutations of $N$ objects, $\pi \in \kS_N$ (describing a complete
assignment of blue and red points), and the cost function is
$H_J(\pi) = \sum_{i=1}^N J_{i \pi(i)}$.

A \emph{random assignment problem} is the datum of a probability
measure $\mu_N(J)$ on the possible instances of the problem.  The
interest is in the determination of the statistical properties of this
problem w.r.t. the measure under analysis, and in particular the
statistical properties of the optimal configuration.  The problem can
be formulated as the zero-temperature limit of the statistical
mechanics properties of a disordered system, where the disorder is the
instance $J$, the dynamical variables are encoded by $\pi$, and the
Hamiltonian is the cost function $H_J(\pi)$.


The case of random assignment problem in which the entries $J_{ij}$
are random  i.i.d.\ variables,  presented already in~\cite{Orland1985}, has been
solved at first, in a seminal paper by Parisi and
M\'ezard~\cite{Mezard1985}, through the replica trick and afterwards 
by the Cavity Equations~\cite{Mezard1986a} (see
also~\cite{Caracciolo:168} for a recent generalization of those
results also at finite system size).  The Parisi-M\'ezard solution
also leads to the striking prediction that, calling $\pio(J)$ the optimal matching for the
instance $J$ and $\Ho(J)=H_J(\pio(J))$ its optimal cost, the average
over all instances of $\Ho$, for $N$ large, tends to $\frac{\pi^2}{6}$.

This problem is simpler than a spin glass, as the
determination of the optimal configuration is feasible in polynomial
time (for example, through the celebrated Hungarian
Algorithm~\cite{lovasz2009matching}), however it remains non-trivial, and
the thermodynamics is replica-symmetric, although the stability of the
RS phase becomes marginal in the zero-temperature limit (as evinced,
again, via the analysis of the cavity equations).  A number of
distinct other approaches to the model, and the solution of the Parisi conjecture~\cite{Parisi-conj} for the behaviour at finite size in the case of an exponential distribution of the random costs, which is the simplest case of a more general conjecture by
Coppersmith and Sorkin~\cite{CopSor}, have appeared later
on~\cite{Aldous2001,Nair2005,Linusson}, and it is fair to say that, up
to date, this model is one of the best-understood and instructive
glassy system.

In analogy with the challenge of understanding spin glasses in finite
dimension, there is an interest towards the study of measures
$\mu_N(J)$ which are induced by some random process in a
$d$-dimensional domain.  For example, the blue and red points could be
drawn randomly in some compact domain $\Omega \subset \bR^d$ (e.g.,
uniformly and independently), with the entry $J_{ij}$ given by some
cost function $c(x_i,y_j)$ where $x_i$ and $y_j$ are the coordinates
in $\Omega$ of the $i$-th blue point, and the $j$-th red point,
respectively.  In the simplest versions of the model, the cost will be
invariant by translations and rotations, so it will just be a function
of the Euclidean distance between the two points~\cite{Mezard1988},
$c(x,y)=c(|x-y|)$.  Scaling and universality arguments suggest to
consider cost functions with a power-law behaviour, that is, for some
parameter $p$, we shall consider the cost function
\begin{equation}
c(x_i,y_j) = \pm |x_i - y_j|^p \, .
\end{equation}
Moreover, the properties of the solution of the assignment problem
depend strongly on the monotonicity and concavity of the cost
function, and power-law costs span a variety of combinations of such
behaviours as $p$ varies in $\mathbb{R}$ (and the sign is considered).
This observation suggests that models with different values of $p$,
and different signs of the cost function, are potentially in different universality classes,
so that it is desirable to perform a study of the model both at
arbitrary $d$ and at arbitrary $p$ (and sign).

For a cost function $c(x,y) = s\, |x-y|^p$, with $s=\pm 1$, we say that
we are in the \emph{attractive} case if $sp>0$ (that is, the cost
increases as the distance increases), and in the \emph{repulsive} case
if $sp<0$ (that is, the cost decreases as the distance increases). The
limit $p \to 0$ also makes sense, as the cost reads (with $s=\pm 1$)
$c(x)=s |x|^p=s (1+p \ln|x|+\mathcal{O}(p^2))$, but, as 
$\pio(\{J_{ij}\})=\pio(\{\kappa J_{ij}+\lambda_i+\mu_j\})$, 
at zero temperature this is
equivalent to the cost
$c(x)=s' \ln|x|+\mathcal{O}(p)$, where $s'=s$ for the limit 
$p \searrow 0$ and $s'=-s$ for the limit 
$p \nearrow 0$.

For the $p>1$ attractive case (i.e.\ monotone increasing and convex cost
function), if the subset $\Omega \subset \mathbb{R}^d$ is compact, it
is known~\cite{Ajtai} that the average total cost $\mathbb{E}_{N}(H_{\rm opt})$
of the optimal assignment scales with the number of points
$N$ according to
\begin{equation}
    \mathbb{E}_{N}(H_{\rm opt})  \simeq 
    \begin{cases}
        N^{1-\frac{p}{2}} & \hbox{for \, } d=1\\
        N^{1-\frac{p}{2}}\, (\log N)^\frac{p}{2} & \hbox{for \, } d=2\\
        N^{1-\frac{p}{d}} & \hbox{for \, } d>2
    \end{cases}
\end{equation}
where $a(N) \simeq b(N)$ if $\frac{a(N)}{b(N)}$ tends to a non-zero
finite constant for $N \rightarrow \infty$.  In this case a relation
with the classical \emph{optimal transport problem} in the continuum has been
exploited~\cite{Caracciolo:158}, in particular for the case $p=2$
where very detailed results have been obtained~\cite{Caracciolo:162,
  Caracciolo:163, Ambrosio2016}. 

If, at size $N$, the points are sampled uniformly in the domain
$\Omega_N := N^{\frac{1}{d}} \Omega$ (as to keep the average density
of order 1), we just have to scale the result above by the factor
$N^{\frac{p}{d}}$.

Recently also the case in which $\Omega$ is not compact, and the
points are not sampled uniformly, has been
considered~\cite{Bobkov2016, Caracciolo:172, Talagrand2018}.

In the repulsive case $p<0$
the cost function $c$ is still {\em convex}. As far as we know, for this
version of the problem only the case $d=1$ has been studied in detail.
In this case~\cite{Caracciolo:160, Caracciolo:169}:
\begin{equation}
    \mathbb{E}_{N}(H_{\rm opt})  \simeq N \, .
\end{equation}
More precisely, in $d=1$ corrections to the leading order in the
large-$N$ expansion were studied.  Among the results, we could derive
the first finite-size corrections for $p<0$, and an explicit
expression in $N$ and $p$ for
$p>1$ \cite{Caracciolo:177}:
\begin{equation}
    \mathbb{E}_{N}(H_{\rm opt}) = 
    \begin{cases}
\displaystyle{
        N \frac{\,\Gamma\left(1+\frac{p}{2}\right)}{p+1}
        \frac{\,\Gamma\left(N+1\right)}{\,\Gamma\left(N+1+\frac{p}{2}\right)}
        \simeq \frac{\,\Gamma\left(1+\frac{p}{2}\right)}{p+1}
        N^{1-\frac{p}{2}}
}
& \hbox{for \, } p>1\\
\rule{0pt}{16pt}%
\displaystyle{
        N \frac{1}{2^p} \left[1 + \frac{1}{3N} \frac{p(p-2)
            (p-4)}{p-3}+ o\left(\frac{1}{N}\right) \right] 
}
& 
\hbox{for \, } p<0\, .
    \end{cases} \label{results}
\end{equation}
The attractive case $p<0$ has not been studied so far, but is possibly
of interest. However, it is easily seen that in this case the quantity 
$\mathbb{E}_{N}(H_{\rm opt})$ makes sense only when $|p|<d$, otherwise
one just gets $\mathbb{E}_{N}(H_{\rm opt})=-\infty$ already from the
singular portion of the measure associated to the rare instances in
which a red and a blue point almost coincide.

When $0<p<1$, in the attractive case, the cost function $c$ is instead
{\em concave}. Also in this case only the case $d=1$ has been
considered, where it has been shown that the optimal solution is
always {\em non-crossing}~\cite{McCannRobert1999, Caracciolo:159}.  A
matching is non-crossing if each pair of matched points defines an
interval on the line which is either nested to or disjoint from all
the others.

Contrary to the case $p>1$, where in $d=1$ the optimal matching is
completely determined, and to the case $p<0$, where in $d=1$ it is
known that the optimal matching has certain cyclic properties, the
non-crossing property is not sufficient to fully characterize the
optimal assignment; the regime $0<p<1$ is thus much more challenging
to study.

The relevance of non-crossing matching configurations among elementary
units aligned on a line has emerged both in physics and in biology. In
the latter case, this is due to the fact that they appear in the study
of the secondary structure of single stranded DNA and RNA chains in
solution~\cite{Higgs2000}.  These chains tend to fold in a planar
configuration, in which complementary nucleotides are matched, and
planar configurations are exactly described by non-crossing matchings
between nucleotides.  The secondary structure of a RNA strand is
therefore a problem of optimal matching on the line, with the
restriction on the optimal configuration to be
planar~\cite{Orland2002, Vernizzi, Nechaev2013}.  The statistical
physics of the folding process is highly non-trivial and it has been
investigated by many different techniques~\cite{Bundschuh2002,
  Muller2003rna}, also in presence of disorder and in search for
glassy phases~\cite{Nechaev2013, Bundschuh2002, Higgs1996,
  Pagnani2000, Marinari2002}.  So, as a further motivation for the
present work, understanding the statistical properties of the solution
to random Euclidean matching problems with a concave cost function
could yeld results and techniques to better understand these models of
RNA secondary structure.

A summary of the scaling behaviours for the different declinations of
the models is provided in the Conclusion section.

\subsection{Random Assignment Problems studied in this paper}

In this paragraph we describe the problem of $0<p<1$ random assignment
in detail, and fix our notations.

First, we fix $\Omega$ to be a segment (an alternate possible choice,
1-dimensional in spirit, which is not considered here but is
considered, for example, in~\cite{Caracciolo:159, Caracciolo:160,
  Caracciolo:163}, is to take $\Omega$ as a circle, or more generally,
in dimension $d$, to take $\Omega$ as a $d$-dimensional
torus~\cite{Caracciolo:158}).


Notice that in the literature $\Omega$ is typically taken to be
deterministically the unit segment $[0,1]$,
while in our paper we will find easier to work with a segment $[0,L]$,
where $L$ may be a stochastic variable, whose distribution depends on
$N$. We will work in the framework of constant density, that is
$\eee(L) \sim N$, so that, for comparison with the existing
literature, our results should be corrected by multiplying by a factor
of the order $N^{-p}$.

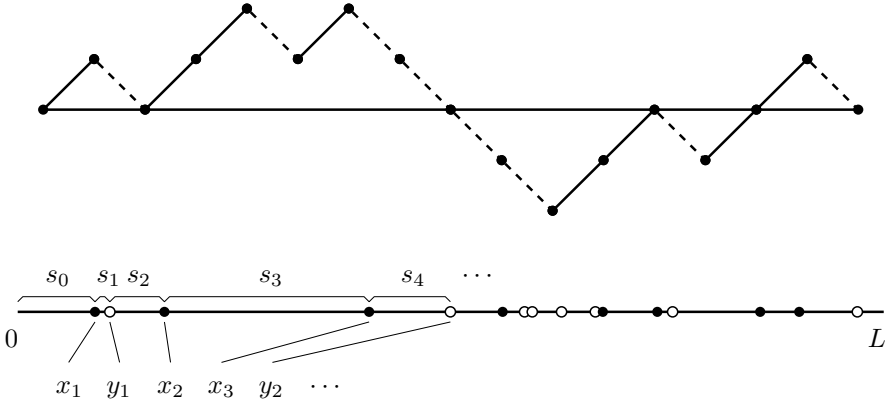
\begin{figure}
\setlength{\unitlength}{19pt}
\[
\begin{picture}(17,5)
    \put(0,0){\includegraphics[scale=.475]{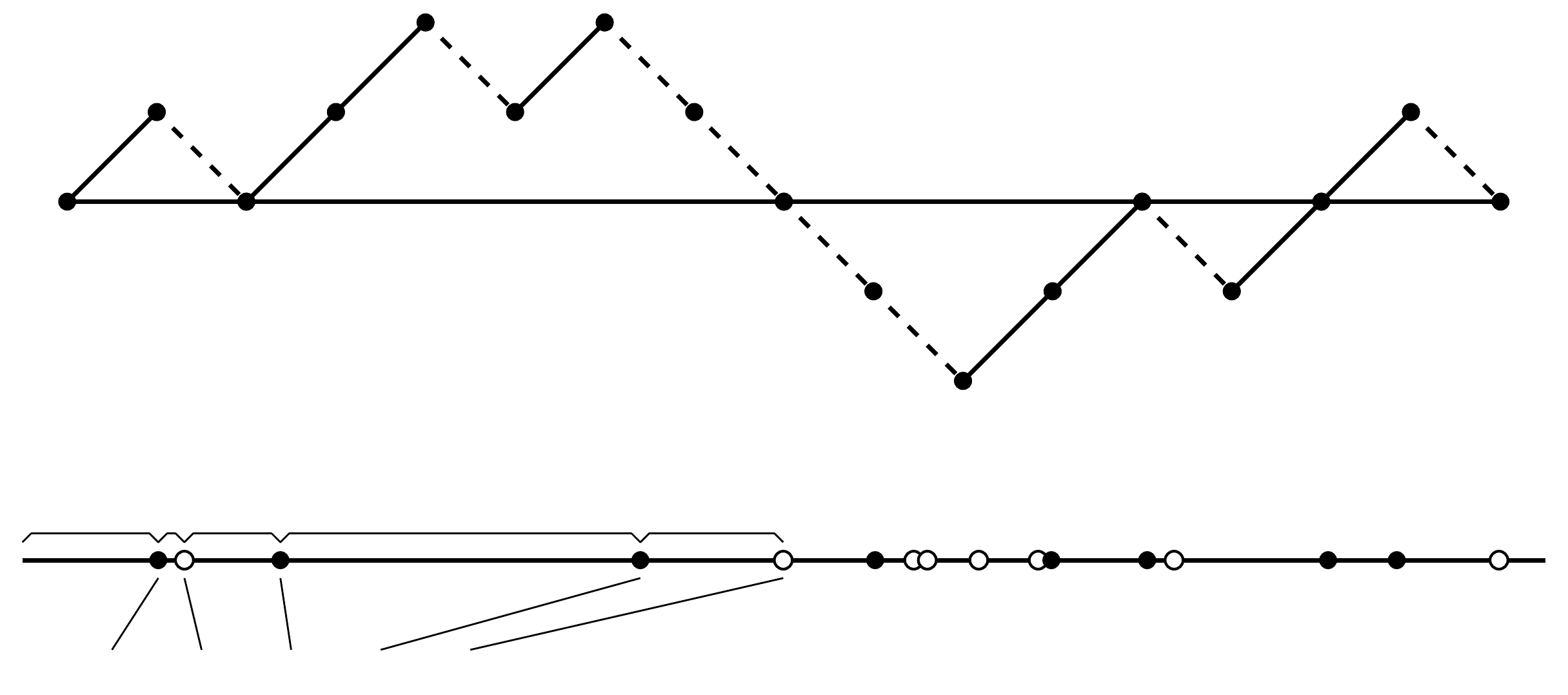}}
\put(1,-.1){$x_1$}
\put(2,-.1){$y_1$}
\put(3,-.1){$x_2$}
\put(4,-.1){$x_3$}
\put(5,-.1){$y_2$}
\put(6,-.1){$\cdots$}
\put(0.8,2.1){$s_0$}
\put(1.8,2.1){$s_1$}
\put(2.4,2.1){$s_2$}
\put(5.0,2.1){$s_3$}
\put(7.8,2.1){$s_4$}
\put(9,2.1){$\cdots$}
\put(0,.8){0}
\put(17,.8){$L$}
\end{picture}
\]
\caption{\label{fig.exam}Example of instance
  $J=(\vx,\vy)=[\vs,\sig]$, in the PPP model.
Bottom: the configuration of points. Top: the Dyck
  bridge associated to $\sig$. Here $N=8$, and $\sig=\{+,-,+,+,-,+,-,-,-,-,+,+,-,+,+,-\}$.}
\end{figure}

In dimension 1 there is a natural ordering of the points, so that we
can encode an instance $J=(\vx,\vy)$ by the ordered lists
$\vx=(x_1,x_2,\ldots,x_N)$, $x_i \leq x_{i+1}$, and
$\vy=(y_1,y_2,\ldots,y_N)$, $y_i \leq y_{i+1}$.  A useful alternate
encoding of the instance is $J=[\vs, \sigma]$, where
$s=(s_0,s_1,\ldots,s_{2N})$, $s_i \in \bR^+$, encodes the distances
between consecutive points (and between the first/last point with the
respective endpoint of the segment $\Omega$) and the vector
$\sigma=(\sig_1,\sig_2,\ldots,\sig_{2N}) \in \{-1,+1\}^{2N}$, with
$\sum_i \sig_i=0$, encodes the sequence of colours of the points (see
Figure \ref{fig.exam}, where the identification $\mathrm{blue}=+1$ and
$\mathrm{red}=-1$ is adopted).  In other words, the partial sums of $\vs$,
i.e.\ $(s_0,s_0+s_1,s_0+s_1+s_2,\ldots,s_0+\cdots+s_{2N-1})$,
constitute the ordered list of $\vx \cup \vy$, and $\sig$ describes
how the elements of $\vx$ and $\vy$ do interlace.
In this notation, the domain of the instance $\Omega=[0,L]$ is
determined by $L=\sum_{i=0}^{2N} s_i$.
Remark that the cardinality of the
space of possible vectors $\sigma$ is just the central binomial,
\begin{equation}
B_N := \binom{2N}{N} 
\ef.
\label{cb}
\end{equation}
For simplicity, in this paper we consider only the non-degenerate
case, in which almost surely all the $s_i$'s are strictly positive,
that is, the values in $\vx \cup \vy$ are all distinct.

In this paper, and more crucially in subsequent work, we shall
consider two families of measures. In all these measures, we have a
factorisation $\mu([\vs,\sigma])=\mu_1(\vs) \mu_2(\sigma)$, and the
measure on $\sigma$ is just the uniform measure.
\begin{description}
    \item[Independent spacing models (ISM).] 
        The measure $\mu(\vs)$ is factorised, and the
        $s_i$'s are i.i.d.\ with some distribution $f(s)$ with support
        on $\bR^+$ (and, for simplicity, say with all moments finite,
        $\int \dx{s} s^k f(s)<\infty$ for all $k$).
        Without loss of generality, we will assume that the average of
        $f(s)$ is $1$, i.e.\ the average of $L$ is $2N+1$.  In
        particular, we will consider:
        \begin{description}
            \item[Uniform spacings (US):] the $s_{i}$'s are
              deterministic, identically equal to~1, and thus $L=2N+1$
              for all instances;
            \item[Exponential spacings (ES):] the $s_{i}$'s are
              i.i.d. with an exponential distribution 
              $f(s) = g_{1}(s)= \exp(-s)$, and thus $L$ concentrates
              on $2N+1$, but has a variance of order $N$.
        \end{description}
    \item[Exchangeable process model (EPM).] This is a generalisation of
        the ISM above,
        but now the $s_i$'s
        are not necessarily i.i.d., they are instead
        \emph{exchangeable variables}, that is, for all $0 \leq i <
        2N$,
        \be
        \mu(s_0,\ldots,s_i,s_{i+1},\ldots,s_{2N}) =
        \mu(s_0,\ldots,s_{i+1},s_{i},\ldots,s_{2N})
        \ef.
        \ee
        In particular, within this class of models we could have that $\mu$
        is supported on the hyper-tetrahedron $\bT_{2N}$ described by $s_i
        \geq 0$, and $L = \sum_i s_i=2N+1$.
        In this paper, we will consider:
        \begin{description}
            \item[Poisson Point Process (PPP):] the $s_i$'s are the spacings among the sorted list of $0$, $2N+1$, and $2N$ uniform random points in the interval $[0,2N+1]$.
        \end{description}
\end{description}

\noindent
Each of these three models has its own motivations.  The PPP case is,
in a sense, the most natural one for what concerns applications and
the comparison with the models in arbitrary dimension~$d$. Implicitly,
it is the one described in the introduction.  
The ES case is useful due to a strong relation with the PPP case (see Remark
\ref{rmk.ESvsPPP1} and Lemma \ref{lem.ESvsPPP2} later on in Section
\ref{sec.linkESPPP}). In a sense, it is the ``Poissonisation'' of the
PPP case (where in this case it is the quantity $L$ that has been
``Poissonised'', that is, it is taken stochastic with its most natural
underlying measure, instead of deterministic). The US case will prove
out, in future work, to be the most tractable case for what concerns
lower bounds to the optimal cost.

As all of the measures above are factorized in $\sigma$ and $\vs$, and
the measure over $\sigma$ is uniform,
it is useful to define two shortcuts for two recurrent notions of
average. 
\begin{definition} \label{rem.factorizeAvg}
For any quantity $A(J)=A(\sigma, \vs)$, we denote by
$\overline{A}$ the average of $A$ over $\sigma$
\begin{equation}
                    \overline{A} :=
\eee_{\sigma}(A)
=\frac{1}{B_N}
\sum_{\sigma} A(\sigma, \vs) \, ;
            \end{equation}
This average is independent from the choice of model among the classes
above.
We denote by 
\begin{equation}
\langle A \rangle : =\eee_{\mu(\vs)}(A)
\end{equation} 
the average of $A$
over $\vs$, with its appropriate measure dependence on the choice of
model.
Finally, we denote by $\mathbb{E}_N(A)$ the result of both averages,
in which we stress the dependence from the size parameter $N$ in the
measure, that is
\be
\mathbb{E}_N(A):=\eee_{\sig,\mu(\vs)}(A)=
\overline{\langle A \rangle}=
\langle \overline{A} \rangle
\ef.
\ee
\end{definition}
For a given instance, parametrised as $J=(\vx,\vy)$, or as
$J=[\vs,\sig]$, (and in which the cost function also has an implicit
dependence from the exponent $p$), call $\pio$ one optimal
configuration, and $\Ho(J)=H_J(\pio)$.


\bigskip

In this paper, we will introduce the notion of \emph{Dyck matching}
$\pid$ of an instance $J$ and we will compute its average cost
$\Hd:=\eee_N(H_J(\pid))$ for the measures ES and PPP (with a brief
discussion on the US case).

In particular we prove the following theorem:
\begin{theorem} \label{thm.main}
For the three measures ES, US and PPP,
    let $\mathbb{E}_{N}(H_{\rm Dyck})$ denote the average cost of the
    Dyck matchings. Then
    \be
    \eee_{N}(H_{\rm Dyck} ) \simeq \left\{
        \begin{array}{ll}
            N & 0 \leq p < \frac{1}{2} \\
            N \ln N & p = \frac{1}{2} \\
            N^{\frac{1}{2}+p} & \frac{1}{2} < p \leq 1
        \end{array}
    \right.
    \ee
    where $a(N) \simeq b(N)$ if $\frac{a(N)}{b(N)}$ tends to a finite,
    non-zero constant when $N \rightarrow \infty$.
\end{theorem}
This theorem follows directly from the combination of suitable lemmas,
namely Proposition~\ref{prop.ES}, Proposition~\ref{prop.US} and
Corollary~\ref{cor.PPP} appearing later on. In fact, our results are
more precise than what is stated in the theorem above (we describe the
first two orders in a series expansion for large $N$, including
formulas for the associated multiplicative constants), details are
given in the forementioned propositions.

The average cost of Dyck matchings provides an upper bound on the
average cost of the optimal solution; numerical simulations for the
PPP measure, described in Section \ref{sec.numerics},
suggest the following conjecture, that we leave for future investigations:
\begin{conj}
\label{conj.main}
For the three measures ES, US and PPP, and all $0<p<1$,
    \begin{equation}
        \lim_{N\rightarrow\infty} \frac{\mathbb{E}_{N}(H_{\rm opt})}{\mathbb{E}_{N}(H_{\rm Dyck})} = k_{p}   \, ,
    \end{equation}
    with $0 < k_{p} < 1$. 
\end{conj}

\section{Basic facts}

Before starting our main proof, let us introduce some more notations,
and state some basic properties of the optimal solution.

\subsection{Basic properties of the optimal matching}
\label{ssec.basicprops}


A {\em Dyck path of semi-length $N$} is a lattice path from $(0,0)$ to
$(2 N,0)$ consisting of $N$ `up' steps (i.e.\ steps of the form
$(1,1)$) and $N$ `down' steps (i.e., steps $(1,-1)$), which never goes
below the $x$-axis. There are $C_N$
Dyck path of semi-length $N$, where
\begin{equation}
C_N = \binom{2 N}{N} - \binom{2 N}{N +1} = \frac{1}{N+1} \binom{2N}{N}
\end{equation}
are the {\em Catalan numbers}. Therefore the generating function for the Dyck paths is
\begin{equation}
C(z) := \sum_{k\ge 0} C_k z^k =
\frac{1 - \sqrt{1- 4z}}{2z} = \frac{2}{1 + \sqrt{1- 4z}}
\end{equation}
The historical name `Dyck path' is somewhat misleading, as it leaves
us with no natural name for the most obvious notion, that is, the
walks of length $N$ with steps in $\{(1,1),(1,-1)\}$. With analogy
with the theory of Brownian motion (which relates to lattice walks via
the Donsker's theorem)~\cite{PitYorguide},
we will define four types of paths, namely walks,
meanders, bridges and excursions, according to the following table:
\begin{center}
\begin{tabular}{r|cc|}
           & $y(2N)=0$ & $y(x)\geq 0 \; \forall x$ \\
\hline
walk       & no        & no         \\ 
meander    & no        & yes        \\ 
bridge     & yes       & no         \\ 
excursion  & yes       & yes        \\ 
\hline
\end{tabular}
\end{center}
(of course, by `no' we mean `not necessarily'). Thus, in fact, the
`paths' are the most constrained family of walks, that is the excursions.

In general a Dyck path (i.e., a Dyck excursion) can touch the $x$-axis
several times. We shall call an {\em irreducible Dyck excursion} a
Dyck path which touches the $x$-axis only at the two endpoints. It is
trivially seen that the generating function for the irreducible Dyck
excursions is simply $z\, C(z)$.
As we said above a {\em Dyck bridge} is a walk made with the same kind
of steps of Dyck paths, but without the restriction of remaining in
the upper half-plane, and which returns to the $x$-axis. The
generating function for the Dyck bridges is
\begin{equation}
B(z) := \frac{1}{1-2 z\,C(z)} = \frac{1}{\sqrt{1-4z}} = \sum_{k\ge 0} B_k z^k
\end{equation} 
with $B_k$ the central binomials~(\ref{cb}), and $k$ is the
semi-length of the bridge (just like excursions, all Dyck bridges must
have even length).  The factor $2$ in the functional form of $B(z)$ in
terms of $C(z)$ enters because a bridge is a concatenation of
irreducible excursions, each of which can be in the upper- or the
lower-half-plane.

Now, it is clear that each configuration $\sigma$ corresponds uniquely
to a Dyck bridge of semi-length $N$, with $\sigma_i = +1$ or $-1$ if the
$i$-th step of the walk is an up or down step, respectively.

In a Dyck walk we shall call $(\kb(i),\hb(i))$ the two coordinates
of the mid-point of the $i$-th ascending step of the walk (minus
$(\smfrac{1}{2},\smfrac{1}{2})$, in order to have integer coordinates
and enlighten the notation), and call $(\kr(i),\hr(i))$ the
coordinates of the mid-point of the $i$-th descending step (again,
minus $(\smfrac{1}{2},\smfrac{1}{2})$).  For $e=(i,j)$ an edge of a
matching $\pi$, call $\|e\|=\kb(i)-\kr(j)$ the horizontal distance on
the walk, and $|e|=|x_i-y_j|$ the Euclidean distance on the domain
segment.

For a given Dyck bridge $\sig$, we say that $\pi \in \kS_N$ is
\emph{non-crossing} if, for every pair of distinct edges
$e_1=(i_1,j_1)$ and $e_2=(i_2,j_2)$ in $\pi$, we do not have the
pattern $\kb(i_1)<\kb(i_2)<\kr(j_1)<\kr(j_2)$, or the analogous
patterns with $\kb(i_1) \leftrightarrow \kr(j_1)$, or $\kb(i_2)
\leftrightarrow \kr(j_2)$, or $(\cdot)_1 \leftrightarrow (\cdot)_2$.
The name comes from the fact that, if we represent a matching $\pi$ as
a diagram consisting of the domain segment $[0,L]$, and the set of $N$
semicircles above this segment connecting the $x_i$'s to the
$y_{\pi(i)}$'s (as in the bottom part of Figure \ref{fig.piDyck}),
then these semicircles do not intersect if and only if $\pi$ is
non-crossing. Note that, although these semicircles are drawn on the
full $[\vs,\sigma]$ instance, the notion of $\pi$ being non-crossing
only uses the vector $\sigma$.

For a given Dyck bridge $\sig$, we say that $\pi \in \kS_N$ is
\emph{sliced} if, for every edge $e=(i,j)\in \pi$, we have
$\hb(i)=\hr(j)$.

Two easy lemmas have a crucial role in our analysis.
\begin{lemma}
\label{lem.NoCross}
All the optimal matchings are non-crossing.
\end{lemma}
\pf\ The proof is by absurd. Suppose that $\pi$ is a crossing optimal
matching.  If we have a pattern as
$\kb(i_1)<\kr(j_2)<\kr(j_1)<\kb(i_2)$, then the matching $\pi'$ with
edges $e'_1=(i_1,j_2)$ and $e'_2=(i_2,j_1)$ has $H_J(\pi')<H_J(\pi)$,
because $|e'_1|<|e_1|$ and $|e'_2|<|e_2|$.

If we have a pattern as $\kb(i_1)<\kb(i_2)<\kr(j_1)<\kr(j_2)$, then
again the matching $\pi'$ with edges $e'_1=(i_1,j_2)$ and
$e'_2=(i_2,j_1)$ has $H_J(\pi')<H_J(\pi)$, although this holds for a
more subtle reason. Calling $a=x_2-x_1$, $b=y_1-x_2$ and $c=y_2-y_1$,
we have $|e_1|=a+b$, $|e_2|=b+c$, $|e'_1|=a+b+c$ and $|e'_2|=b$. It is
the case that, for $a,b,c >0$ and $0<p<1$,
\be
(a+b)^p+(b+c)^p > (a+b+c)^p + b^p
\ef.
\ee
A proof of this inequality goes as follows. Call
$
F(a,b,c)=(a+b)^p+(b+c)^p-(a+b+c)^p-b^p
$.
We have
$F(a,b,0)=0$, and
\be
\frac{1}{p}
\deenne{c}{{}}
F(a,b,c)
=
\frac{1}{(b+c)^{1-p}}-\frac{1}{(a+b+c)^{1-p}}
>0
\ef.
\ee
All the other possible crossing patterns are in the first or the
second of the forms discussed above, up to trivial symmetries.  \qed

\bigskip
\noindent
Non-crossing properties for assignment and optimal transport problems
with concave cost functions were studied in the continuum case in
\cite{McCannRobert1999} and in the discrete one in
\cite{Caracciolo:159}.

\begin{lemma}
\label{lem.sliced}
All the optimal matchings are sliced.
\end{lemma}
\pf\ The proof is by absurd.  Suppose that $\pi$ is a non-sliced
optimal matching.  If we have $(i,j) \in \pi$ with $\hb(i) \neq
\hr(j)$, say $\hb(i) -\hr(j) = \delta h \neq 0$, we have that the
point $x_i$ is matched to $y_j$, and that, between $x_i$ and $y_j$,
there are
$n_{\rm blue}$ and $n_{\rm red}$ blue and red points, respectively,
with $n_{\rm blue}-n_{\rm red} = -\delta h \neq 0$. So there must be
at least $|\delta h|$ points inside the interval $(x_i,y_j)$ which are
matched to points outside this interval, and thus, together with
$(i,j)$, constitute pairs of crossing edges.  So, by
Lemma~\ref{lem.NoCross}, $\pi$ cannot be optimal.  \qed

\vspace{0.1cm}
The slicing of optimal assignments was studied in \cite{delonLocalMatchingIndicators2012}.

\subsection{Reducing the PPP model to the ES model}
\label{sec.linkESPPP}

Theorem \ref{thm.main} (and, hopefully, Conjecture \ref{conj.main}),
in principle, shall be proven for three different models, Uniform
Spacings (US), Exponential Spacings (ES) and the Poisson Point Process
(PPP). However, as we anticipated, the PPP case is a minor variant of
ES. In this section we give a precise account of this fact.

The starting point is a relation between the two measures:
\begin{remark}
\label{rmk.ESvsPPP1}
We can sample a pair $[{\vs}\,',\sig]$ with the measure $\mu^{\rm ES}_N$
by sampling a pair $[\vs,\sig]$ with the measure $\mu^{\rm PPP}_N$, a
value
$L \in \bR^+$ with the measure 
$g_{2N+1}(L)=\frac{L^{2N}}{(2N)!} \exp(-L) \ddx{L}$, and then defining 
$s'_i = s_i \frac{L}{2N+1}$.
\end{remark}
Indeed, the measure on $\vs$ in the PPP can be seen as the measure over independent exponential variables conditioned to the value of the sum; 
thus, the procedure leads at sight to a measure over $\vs\,'$ which is unbiased within vectors $\vs\,'$ with the same value of $L=\sum_i s'_i$. 
Then, from the independence of the spacings in the ES model we easily conclude that the distribution of $L$ must be exactly $g_1^{\ast (2N+1)}(L)$, i.e. the convolution of $2N+1$ exponential distributions.

More precisely, for $k$ an integer, define the Gamma measure
\be
g_{k}(x):=\frac{x^{k-1}}{\Gamma(k)} e^{-x}=g_1^{\ast k}(x)
\ef.
\ee
We use the same notation for its analytic continuation to $k$ real
positive.

We introduce (the analytic continuation of) the rising factorial
(following the notation due to D.~Knuth~\cite{ConcreteMath}): 
\be
n^{\overline{p}}:= \frac{\Gamma(n+p)}{\Gamma(n)}
=\int \dx{x} g_n(x) x^p
\ef.
\ee
This choice of notation is motivated by the fact that $n^{\overline{1}}=n^1=n$,
and that, for $n \gg p^2$, $n^{\overline{p}}=n^p (1+\cO(p^2/n))$. More precisely
we have:
\begin{lemma}
\label{lem.boundsXP}
For $0 \leq p \leq 1$ and $n>1$
\be
(n+p-1)^p \leq n^{\overline{p}} \leq n^p
\ef.
\ee
\end{lemma}
\pf\ It is well known that the Gamma function is logarithmically
convex~\cite{ArtinBook}. In particular, for any 
$0\leq p \leq 1$,
\be
\ln \Gamma(n+p) \leq (1-p) \ln \Gamma(n) + p \ln \Gamma(n+1)
= \ln \Gamma(n) + p \ln n
\ef,
\ee
that is
\be
n^{\overline{p}}= \exp( \ln \Gamma(n+p) - \ln \Gamma(n) ) \leq \exp(p \ln n) =
n^p
\ef.
\ee
Analogously, we have
\be
\ln \Gamma(n) \leq (1-p) \ln \Gamma(n+p) + p \ln \Gamma(n+p-1)
= \ln \Gamma(n+p) - p \ln (n+p-1)
\ef,
\ee
that is
\be
n^{\overline{p}}= \exp( \ln \Gamma(n+p) - \ln \Gamma(n) ) \geq \exp(p \ln (n+p-1)) =
(n+p-1)^p
\ef.
\ee
\qed

Summing up, Remark~\ref{rmk.ESvsPPP1} and Lemma~\ref{lem.boundsXP} allow to prove that
\begin{lemma}
\label{lem.ESvsPPP2}
The following inequalities hold:
\be
\left(\frac{2N}{2N+1}\right)^{p(1-p)}
\eee(\Ho^{\rm ES})
\leq
\eee(\Ho^{\rm PPP}) 
\leq 
\eee(\Ho^{\rm ES})
\ef.
\ee
Also the corresponding inequalities with $\Ho$ replaced by $\Hd$ do
hold, as well as for any other quantity $H(\pi^*(J))$, whenever
$\pi^*$ is some matching determined by the instance, and 
invariant under scaling of the instance, that is
$\pi^*[\vs,\sig]=\pi^*[\lam\vs,\sig]$.
\end{lemma}
\pf\ For compactness of notation, we do the proof only for the $H_{\rm
  opt}$ case, but the generalisation is straightforward.  Of course,
$H_{[\lam \vs,\sig]}(\pi)=\lam^p H_{[\vs,\sig]}(\pi)$
for all instances $[\vs, \sig]$, all configurations $\pi$, and all
scaling factors $\lam > 0$ (so as
$\pi^*[\vs,\sig]=\pi^*[\lam\vs,\sig]$, it follows that
$H_{[\lam \vs,\sig]}(\pi^*[\lam \vs,\sig])=\lam^p
H_{[\vs,\sig]}(\pi^*[\vs,\sig])$).
In particular,
$\Ho([\lam \vs,\sig])= \lam^p \Ho([\vs,\sig])$.
In light of Remark \ref{rmk.ESvsPPP1}, we can describe the
average over $\mu_{\rm ES}[\vs,\sig]$ in terms of an average over
$\mu_{\rm PPP}[\vs,\sig]$, and over $g_{2N+1}(L)$. More precisely
\be
\begin{split}
\eee(\Ho^{\rm ES})
&=
\int \dx{\mu_{\rm ES}[\vs\,',\sig]}
\Ho([\vs\,',\sig]) 
\\
&=
\int \dx{\mu_{\rm PPP}[\vs,\sig]}
\int \dx{L} g_{2N+1}(L) 
\left( \frac{L}{2N+1} \right)^p
\Ho([\vs,\sig])
\\
&= 
\eee(\Ho^{\rm PPP}) 
\int \dx{L} g_{2N+1}(L) \left( \frac{L}{2N+1} \right)^p
\\
&= 
\eee(\Ho^{\rm PPP}) \;
\frac{(2N+1)^{\overline{p}}}{(2N+1)^p}
\ef.
\end{split}
\ee
The upper bound follows directly from Lemma~\ref{lem.boundsXP}.
The lower bound follows as:
\begin{equation}
    \frac{(2N+1)^{\overline{p}}}{(2N+1)^{p} } \geq \left(1-\frac{1-p}{2N+1}\right)^{p} \geq \left(1-\frac{1}{2N+1}\right)^{p(1-p)}    \, 
\end{equation}
where first one uses Lemma~\ref{lem.boundsXP}, then the inequality
$(1-q \eps)^p \geq (1-\eps)^{pq}$ (valid for $\eps, p, q \in [0,1]$),
with $\eps = \frac{1}{2N+1}$ and $q=1-p$.  \qed

In light of this lemma, it is sufficient to consider
Theorem~\ref{thm.main} (and Conjecture \ref{conj.main}) only for the
US and ES models.

The precise statement of our conclusions is the following:
\begin{corollary}   
\label{cor.PPP}
    \begin{equation}
        \begin{split}
\mathbb{E}_{N}^{PPP}(H_{\rm Dyck}) = \mathbb{E}_{N}^{ES}(H_{\rm Dyck})
\left(1 +\mathcal{O}\left(N^{-1}\right)\right)
\ef.
        \end{split}
    \end{equation}
\end{corollary}
The relevance of this statement lies in the fact that, in the
forthcoming equation~(\ref{eq.IS}), we provide an expansion for
$\mathbb{E}_{N}^{ES}(H_{\rm Dyck})$ in which corrections of relative
order $1/N$ appear as the third term in the expansion (and we provide
explicit results only for the first two terms). As a result, the very
same conclusions that we have for the ES model do apply verbatim to
the PPP model.


\section{The Dyck matching}
\label{sec.UB}

\begin{figure}
\setlength{\unitlength}{19pt}
\[
\begin{picture}(17,5)
\put(0,0){\includegraphics[scale=.475]{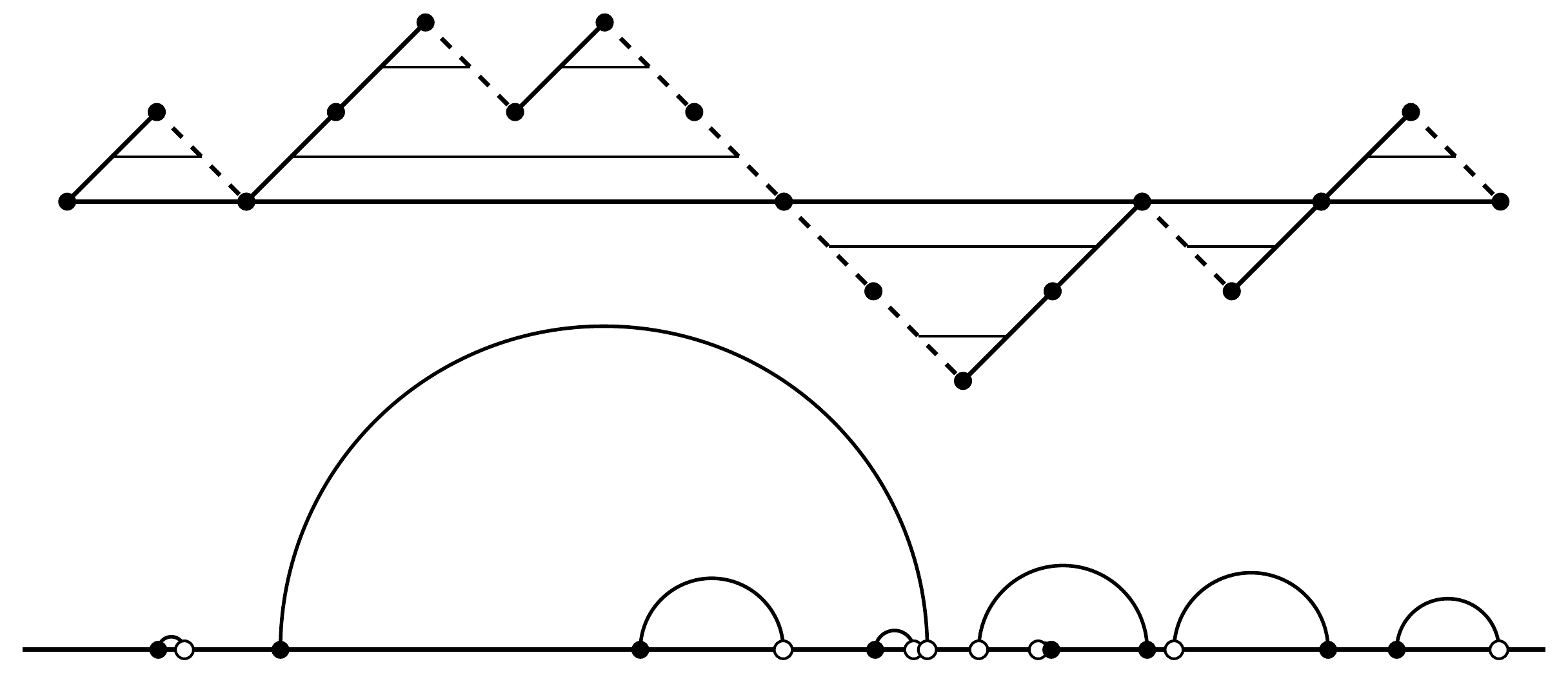}}
\end{picture}
\]
\caption{\label{fig.piDyck}The Dyck matching $\pid$ associated to the
  Dyck bridge $\sig$ in the example of Figure~\ref{fig.exam}.}
\end{figure}

For every instance $[\vs,\sig]$, there is a special matching, that we
call $\pid$, which is sliced and non-crossing for $\sig$.
This is the matching obtained by the canonical pairing of up- and
down-steps within every excursion of the Dyck bridge
(see Figure~\ref{fig.piDyck}).
In analogy with our notation $\Ho(J)=H_J(\pio)$, let us introduce the
shortcut $\Hd(J)=H_J(\pid)$.

\begin{remark}\label{rem.DyckAvg}
        The Dyck matching $\pi_{\rm Dyck} $ is determined by the order
        of the colours of the points, $\sigma$. In particular it does
        not depend on the actual spacings between them, $\vs$, and it
        does not depend on the cost exponent $p$.
\end{remark}
Remark~\ref{rem.DyckAvg} is a crucial fact that makes possible the
evaluation of the statistical properties of $\pid$, with a moderate
effort. In particular, it will lead to the main result of this paper:
\begin{proposition}
\label{prop.ES}
For all independent spacing models
\begin{equation} 
\label{eq.IS}
    \begin{split}
        \mathbb{E}_{N} (H_{\rm Dyck})\sim  \, 
            \begin{cases}
                \frac{A_{p}}{2} N + 
                \frac{2^{p} \Gamma(p-\frac{1}{2})}{4 \Gamma(p+1)}
                N^{\frac{1}{2}+p} 
+ \mathcal{O}(1) 
&\, p < \frac{1}{2}\\ 
                \frac{1}{\sqrt{2\pi}} N \log{N} + 
\left( \frac{A_{\ast}}{2} + A'_{\ast} \right)
N + 
\mathcal{O}(\log N) 
&\, p = \frac{1}{2}  \, .\\
                \frac{2^{p} \Gamma(p-\frac{1}{2})}{4 \Gamma(p+1)}
                N^{\frac{1}{2}+p} 
+ \frac{A_{p}}{2} N 
+ \mathcal{O}(N^{-\frac{1}{2}+p}) &\, p > \frac{1}{2}\\ 
            \end{cases}
    \end{split}
\end{equation}
where $A_{p}$ and $A_{*}$ are model-dependent quantities, which are
not larger than
\begin{align}
A^{\rm max}_{p} 
&:=\frac{2^{p+1}}{(1-2p)\Gamma\left(1-p\right)}
\ef,
&
A^{\rm max}_{*}
&=
\sqrt{\frac{2}{\pi}} \left( \log 2 + \gamma_{E} \right)\, .
\end{align}
and 
\be
A'_{\ast} = \frac{\gamma_{E} + 2\log2 -2}{\sqrt{2\pi}} \, .
\ee
In particular, for the ES model
\begin{align}
A^{\rm ES}_{p} &=
\frac{\Gamma\left(\frac{1}{2}-p\right)\,\Gamma\left(p+1\right)}{2^{p-1}\,
  \sqrt{\pi}\,\Gamma\left(2-p\right)}
&
A^{\rm ES}_{\ast} &= \sqrt{\frac{2}{\pi}} \, \left( 5 \log 2 + \gamma_{E} - 4 \right) \, .
\end{align}
\end{proposition}

The remaining of this section is devoted to the proof of this
proposition.
First, we factorize the average over the instance
$J=[\vs,\sigma]$ in two independent averages, the average over
$\sigma$ and the one over $\vs$ (see
Definition~\ref{rem.factorizeAvg}):
\begin{equation}
    \begin{split}
\mathbb{E}_{N}(H_{\rm Dyck}) &= \mathbb{E} 
\left(\sum_{i=1}^{N} J_{i, \pi_{\rm Dyck} (i)}\right) = 
\overline{\sum_{i=1}^{N} \langle J_{i, \pi_{\rm Dyck} (i)} \rangle }
\\ 
&=
B_{N}^{-1} \sum_{\sigma} \sum_{i=1}^{N} 
\left\langle | \kb(i) - \kr(\pi^{\sigma}_{\rm Dyck}(i)) |^p \right\rangle \, ,
    \end{split} 
\end{equation}
where in the last line we emphasize that $\pi_{\rm Dyck}$ depends only
on $\sigma$, as stated in Remark~\ref{rem.DyckAvg}, and we adopted again
the shortcut $B_{N} = \binom{2N}{N}$ for the total number of
configurations $\sigma$.

Due to the fact that the spacings $s_i$ are independent, the quantity
appearing above,
$\langle | \kb(i) - \kr(\pi^{\sigma}_{\rm Dyck}(i)) |^p \rangle$,
only depends on the length $\|e\|=2k+1$ of the corresponding link
$e=(i, \pid(i))$, via
the formula
\begin{equation}
  S_{k}^{(p)} := 
\left\langle | \kb(i) - \kr(\pi^{\sigma}_{\rm Dyck}(i)) |^p
\right\rangle
=
\bigg\langle \bigg(\sum_{j=0}^{2k} s_{j}\bigg)^p \bigg\rangle
\ef,
\end{equation}
where the $s_j$'s are i.i.d.\ variables sampled with the distribution
$f(s)$ (that is, in the ES model, i.i.d.\ exponential random
variables).

Then, as a general paradigm for observables of the form 
$\overline{\sum_{e \in \pi}\left\langle F(|e|)\right\rangle}$,
we rewrite the sum over all possible $\sigma$
and over all links $e$ of a given matching $\pi=\pid(\sigma)$ as a sum
over the forementioned parameter $k$, with a suitable combinatorial
factor $v_{N,k}$:
\begin{align}
\overline{\sum_{e \in \pi}\left\langle F(|e|)\right\rangle}
&=
B_{N}^{-1} \sum_{k=0}^{N-1} v_{N,k}
\bigg\langle F\bigg(\sum_{j=0}^{2k} s_{j}\bigg) \bigg\rangle
\\
v_{N,k} &=
\sum_{\sigma}
\sum_{e \in \pid(\sigma)}
\delta_{\|e\|,2k+1}
\ef.
\end{align}
(Note that $\sum_k v_{N,k}=N B_N=2 \binom{2N-1}{N}$).
In particular, in our case,
\begin{equation}
\label{eq.sumCost}
    \begin{split}
\mathbb{E}_{N}(H_{\rm Dyck}) &= 
B_{N}^{-1} \sum_{k=0}^{N-1} v_{N,k}
\, S_{k}^{(p)}
\ef.
    \end{split}
\end{equation}
So we face two separate problems: (1) determining the combinatorial
coefficients $v_{N,k}$, which are `universal' (i.e., the same for all
independent-spacing models, for all cost exponents $p$ and for all
observables $F$ as above); (2) determining the quantities
$S_{k}^{(p)}$, that is, the average over the Euclidean length $|e|$ of
the link (which depends from the function $f(s)$ that defines the
independent-spacing model, and from the exponent $p$).

For what concerns $S_{k}^{(p)}$, this can be computed exactly both in
the US and ES cases: in the US case $S_{k}^{(p)} = (2k+1)^p$, as in
fact $|e| = 2k+1 = \|e\|$ deterministically, while in the ES case
$S_{k}^{(p)} = \Gamma(2k+1+p)/\Gamma(2k+1)$.  More generally,
for any model with independent spacings we would have that
$S_{k}^{(p)} = \int \dx{x} x^p f^{\ast 2k+1}(x)$ that is, the sum of
$2k+1$ i.i.d.\ random variables is distributed as the $(2k+1)$-fold
convolution of the single-variable probability distribution. For the
ES case this is exactly the Gamma distribution $g_{2k+1}(s)$.  Up to
this point, we could have also evaluated the analogous quantity for
the PPP model, although with a bigger effort (but, from Section
\ref{sec.linkESPPP}, we know that this is not necessary).

For what concerns $v_{N,k}$, in Appendix~\ref{ap.vnk} we prove that
\begin{equation} 
\label{eq:v}
    \begin{split}
        v_{N,k} = C_{k} \left[ 4^{N-k-1} + \frac{N-k}{2} B_{N-k}
          \right] 
=: C_{k} V_{N-k-1}
\ef.
    \end{split}
\end{equation}
In particular, the simple expression for $V_{N-k-1}$ gives in a
straightforward way
\begin{equation} \label{eq:V}
    \begin{split}
        V(z) := \sum_{j=0}^{\infty} V_{j} z^{j} = (1-4z)^{-1} + (1-4z)^{-\frac{3}{2}}\, .
    \end{split}
\end{equation}
We pause to study the distribution of the lengths $\|e\|$ of links in $\pid$, that is,
the normalised distribution (in $k$), with parameter $N$, given by the
expression $v_{N,k}/(N B_N)$.
It is known that planar secondary structures have a universal
behaviour for the tail of such a distribution, with exponent
$-\frac{3}{2}$. Indeed, by performing a large $N$ expansion at fixed
$k$, and then studying the large $k$ behaviour, one has
\begin{equation}
    \begin{split}
        \frac{v_{N,k}}{N B_N} &= \frac{C_k}{N B_N} \Bigl[ 4^{N-k-1} +\frac{N-k}{2}B_{N-k}\Bigr] \\
      &\overset{N\rightarrow\infty}{\approx} 2\, C_k\, 4^{-k}
      \overset{k\rightarrow\infty}{\approx} \sqrt{\frac{2}{\pi}} \, k^{-\frac{3}{2}} \, ,
    \end{split} 
\end{equation}
reproducing the known behaviour.


Equation~(\ref{eq.sumCost}), with the help of~(\ref{eq:v}) and
(\ref{eq:V}), can be used to relate the generating functions
\begin{align} \label{lemma:EVS}
E(z;p) &:= \sum_{N=0}^{\infty} B_{N}\mathbb{E}_{N}(H_{\rm Dyck})\, z^{N} \\
S(z;p) &:= \sum_{k=0}^{\infty} C_{k}\, S_{k}^{(p)}\, z^{k} \, .
\end{align}
\begin{lemma}
\begin{equation}
E(z;p) = z\, V(z)\, S(z;p)
\end{equation}
\end{lemma}
\pf
\begin{equation}
    \begin{split}
        E(z;p) &:= \sum_{N=0}^{\infty} B_{N}\mathbb{E}_{N}(H_{\rm Dyck}) z^{N} \\
             &= \sum_{N=0}^{\infty} \sum_{k=0}^{N-1} C_{k} V_{N-k-1} S_{k}^{(p)} z^{N} \\
             &= 
\bigg(
\sum_{k=0}^{\infty} C_{k}S_{k}^{(p)}z^{k} 
\bigg)
\bigg(
\sum_{N=k+1}^{\infty} V_{N-k-1} z^{N-k} 
\bigg)
\\
             &= z\, V(z)\, S(z;p)
        \, 
    \end{split}
\end{equation}
\qed

The behaviour at large $k$ of $S_{k}^{(p)}$ is determined by the
theory of large deviations. Said heuristically, the sum of the $2k+1$
i.i.d.\ variables $s_{i}$ concentrates on $2k+1$, with tails which are
sufficiently tamed that the average of $x^p$ is equal to 
$(2k+1)^p (1+\mathcal{O}(k^{-1}))$. That is,
$S_{k}^{(p)} \sim (2k+1)^{p} \sim 2^{p} k^{p}$, and we have
\begin{equation} \label{eq.largeK}
    \begin{split}
        C_{k} S_{k}^{(p)} \sim \frac{2^{p}}{\sqrt{\pi}} 4^{k} k^{p-\frac{3}{2}} \, .
    \end{split}
\end{equation}
We recall a fundamental fact in the theory of generating functions:
the singularities of a generating function determine the asymptotic
behaviour of its coefficients.  In particular, the modulus of the
dominant singularity (that nearest to the origin) determines the
exponential behaviour, and the nature of the singularity determines
the subexponential behaviour (see \cite{flajolet2009analytic} for a
comprehensive treatment of singularity analysis, and
Appendix~\ref{ap:asy} for a summary of results).  This tells us that
we just need an expression for $E(z;p)$ around its dominant
singularity to extract asymptotic information on the total cost,
i.e. we just need to evaluate $S(z;p)$ locally around the dominant
singularity of $E(z;p)$.

First of all, one needs to locate the dominant singularity of $S(z;p)$
and compare it with the $z=\frac{1}{4}$ singularity of $V(z)$.  From
Equation~\ref{eq.largeK}, we find an exponential behaviour $\sim 4^n$
of the coefficient of $S(z;p)$, trivially due to the entropy of Dyck
walks of length $2n$, thus, the singularity must be in $z =
\frac{1}{4}$.  Notice that this agrees with the dominant singularity
of $V(z)$ (which also is, essentially, a generating function of Dyck
walks up to algebraic corrections), so that both generating functions
will combine to give the final average-cost asymptotics.

At the dominant singularity, the power-law behaviour of the
coefficients is given by a generating function of the kind
\begin{equation} \label{eq.ansatz}
    S(z;p) = A_{p} + B_{p} (1-4z)^{g_{p}} + o((1-4z)^{g_{p}}) \, ,
\end{equation}
where $A_{p}$ encodes the regular terms at the singularity, and
$o((1-4z)^{g_{p}})$ accounts for all other singular terms leading to
non trivial subleading corrections (among them one finds power,
logarithms\dots in the variable $1-4z$).

In fact, in such a simple situation as in our case, we expect a more
precise behaviour, $S(z;p) = A_{p} (1+ \mathcal{O}(1-4z)) + B_{p}
(1-4z)^{g_{p}} (1+ \mathcal{O}(1-4z))$, where we have two series of
corrections, in integer powers, associated to the regular and singular
parts of the expansion around the singularity (up to the special
treatment of the degenerate case $g_p \in \mathbb{Z}$).

Notice that $B_{p}$ and $g_{p}$ can be found by computing the
asymptotic behaviour of the coefficients of Equation~\ref{eq.ansatz}
\begin{equation}
    \begin{split}
        S_{k}^{(p)} \sim \frac{B_{p}}{\,\Gamma\left(-g_{p}\right)}
        4^{k} k^{-g_{p}-1} = \frac{2^{p}}{\sqrt{\pi}} 4^{k}
        k^{p-\frac{3}{2}}\, ,
    \end{split}
\end{equation}
giving, by comparison with Equation~\ref{eq.largeK}, 
\begin{align}
        g_{p} &= \frac{1}{2}-p 
&
        B_{p} &= \frac{2^{p} \,\Gamma\left(p-\frac{1}{2}\right)}{\sqrt{\pi}}\, . 
\end{align}
Nothing can be said on the coefficient $A_{p}$ without performing the
exact resummation of the generating function at the singularity
(possibly, after having subtracted a suitable diverging part).

This analysis results in an asymptotic expression for $E(z;p)$:
\begin{equation}\label{eq.asyE}
    \begin{split}
        E(z;p) \sim \frac{1}{4} \left[ A_{p} (1-4z)^{-\frac{3}{2}} +
          \frac{2^{p} \,\Gamma\left(p-\frac{1}{2}\right)}{\sqrt{\pi}}
          (1-4z)^{-(1+p)}  \right] 
    \end{split}
\end{equation}
for $p \neq \frac{1}{2}$, and
\begin{equation}
    \begin{split}
        E\left(z;p=\smfrac{1}{2}\right) &= \frac{1}{4}(1-4z)^{-\frac{3}{2}} \left[ A_{\frac{1}{2}+\epsilon} + \frac{1}{\epsilon}\sqrt{\frac{2}{\pi}} + \sqrt{\frac{2}{\pi}} \log \left( \frac{1}{1-4z} \right) + o(\epsilon) \right] \\
               &= \frac{1}{4}(1-4z)^{-\frac{3}{2}} \left[ A^{*} + \sqrt{\frac{2}{\pi}} \log \left( \frac{1}{1-4z} \right) \right] \\
    \end{split}
\end{equation}
for $p=\frac{1}{2}$, where $\epsilon = p-\frac{1}{2}$.
The hypothesis of $S(z;p)$ being non-singular in $p$ implies that
$A_{p}$ must cancel the $\frac{1}{\epsilon}$ singularity, leaving a
regular part
\begin{equation}
    \begin{split}
        A_{\ast} = \lim_{\epsilon\rightarrow 0} \left[ A_{\frac{1}{2}+\epsilon} + \frac{1}{\epsilon}\sqrt{\frac{2}{\pi}}\right]\,. 
    \end{split}
\label{eq.aast}
\end{equation}
This set of results has remarkable consequences, as it unveils a
certain universality feature for $E(z;p)$.  In fact, for all models in
our large classes, the nature of the dominant singularity of $E(z;p)$
is the same, giving a universal asymptotic scaling in $N$ for the
average cost of Dyck matchings.  Moreover, in the $p\geq\frac{1}{2}$
regime, also the coefficient of the dominant singularity is universal.

We can now expand the generating function using standard techniques
(Appendix~\ref{ap:asy}, \cite{flajolet2009analytic}) and the fact that
$B_{N} \sim \frac{4^{N}}{\sqrt{\pi \, N}}$ , obtaining
\begin{equation}
    \begin{split}
        \mathbb{E}_{N} \sim  \, 
            \begin{cases}
                \frac{A_{p}}{2} N + o(N) &\, p < \frac{1}{2}\\ 
                \frac{1}{\sqrt{2\pi}} N \log{N} + o(N \log{N}) &\, p = \frac{1}{2}  \, .\\
                \frac{2^{p} \Gamma(p-\frac{1}{2})}{4 \Gamma(p+1)} N^{\frac{1}{2}+p} + o(N^{\frac{1}{2}+p}) &\, p > \frac{1}{2}\\ 
            \end{cases}
    \end{split}
\end{equation}
and in fact, more precisely,
\begin{equation}
    \begin{split}
        \mathbb{E}_{N} \sim  \, 
            \begin{cases}
                \frac{A_{p}}{2} N + 
                \frac{2^{p} \Gamma(p-\frac{1}{2})}{4 \Gamma(p+1)}
                N^{\frac{1}{2}+p} 
+ \mathcal{O}(1) 
&\, p < \frac{1}{2}\\ 
                \frac{1}{\sqrt{2\pi}} N \log{N} + 
\left( \frac{A_{\ast}}{2} + A'_{\ast} \right)
N + 
\mathcal{O}(\log N) 
&\, p = \frac{1}{2}  \, .\\
                \frac{2^{p} \Gamma(p-\frac{1}{2})}{4 \Gamma(p+1)}
                N^{\frac{1}{2}+p} 
+ \frac{A_{p}}{2} N 
+ \mathcal{O}(N^{-\frac{1}{2}+p}) &\, p > \frac{1}{2}\\ 
            \end{cases}
    \end{split}
\end{equation}
where the terms of the expansion are just the same for the
$p<\frac{1}{2}$ and $p>\frac{1}{2}$ cases, but have been arranged
differently, in the order of dominant behaviour. The quantity 
$A_{\ast}$ has been defined in (\ref{eq.aast}), for the behaviour of 
$S(z;\smfrac{1}{2})$, while the quantity
\be
A'_{\ast} = \frac{\gamma_{E} + 2\log2 -2}{\sqrt{2\pi}}
\ee
is a further (universal) correction coming from taking into account
how $S(z;\smfrac{1}{2})$ enters in $E(z;\smfrac{1}{2})$, via $V(z)$
(and $\gamma_{E}$ is the Euler--Mascheroni constant).

The formulas above gives the precise asymptotics, up to relative
corrections of the order $N^{-1}$.  As a corollary, we have this very
same behaviour in the PPP model, as, from Lemma~\ref{lem.ESvsPPP2} and
Corollary~\ref{cor.PPP}, we know that also the relative corrections
between ES and PPP models are of the order $N^{-1}$.

For higher-order corrections, one would need to take into account more
subleading terms in Equation~\ref{eq.ansatz}.


For the ES case the resummation of $E(z;p)$ can be performed
analytically by writing the Catalan number in terms of Gamma
functions, namely
\begin{equation}
    \begin{split}
        C_{k} = \frac{4^{k}\,\Gamma\left(k+\frac{1}{2}\right)}{\sqrt{\pi} \,\Gamma(k+2)} \, .
    \end{split}
\end{equation}
giving
\begin{equation}
    \begin{split}
        S(z;p) = \Gamma\left(p+1\right) F\left(\begin{array}{c} \frac{p+1}{2} , \frac{p+2}{2} \\ 2 \end{array} \bigg\rvert \, 4z \right)\, 
    \end{split}
\end{equation}
where $F$ is the ${}_2F_1$ hypergeometric function (a reminder is in
Appendix~\ref{ap:asy}, equation~(\ref{eq:def_hyper})).
This allows for an explicit computation of the two non-universal
quantities in our expansion:
\begin{align}
A^{\rm ES}_{p} &=
\frac{\Gamma\left(\frac{1}{2}-p\right)\,\Gamma\left(p+1\right)}{2^{p-1}\,
  \sqrt{\pi}\,\Gamma\left(2-p\right)}
&
A^{\rm ES}_{\ast} &= \sqrt{\frac{2}{\pi}} \, \left( 5 \log 2 + \gamma_{E} - 4 \right) \, .
\end{align}
note how the $A_{\ast}$ and $A'_{\ast}$ terms involve combinations of
quantities of the same algebraic nature.
See Appendix~\ref{ap:asy} for the details of the derivation.

Similar but more complex resummations seem possible in the independent
spacing case, when the function $f(x)$ is a Gamma distribution,
$f(x)=a g_a(ax)$ for $a \in \mathbb{N}/2$, and $S(z;p)$ is obtained in
terms of generalised hypergeometric functions ${}_{k+1}F_k$. However
no exact resummation seems possible for the US case (which would
require a limit $a \to \infty$ in this procedure).
\section{Numerical results and the average cost of the optimal
  matching}
\label{sec.numerics}

Our main results concern the leading behaviour of the average cost of
the Dyck matching, which, of course, provides an upper bound to the
average cost of the optimal matching.  The explicit investigation of
small-size instances suggests that the optimal matching is often quite
similar to the Dyck matching, in the sense that the symmetric
difference between $\pio$ and $\pid$ typically consists of `few'
cycles, which are `compact', in some sense.  Thus, a natural question
arises: could it be that the large-$N$ average properties of optimal
matchings and Dyck matchings the same? If not, in which respect do
they differ?  In order to try to answer this question, we have
performed numerical simulations by generating random instances with
measure $\mu_{N}^{PPP}$, and we have computed the average cost associated to
the optimal and to the Dyck matching.

\begin{figure}
  \centering
  \begin{center}
      \includegraphics{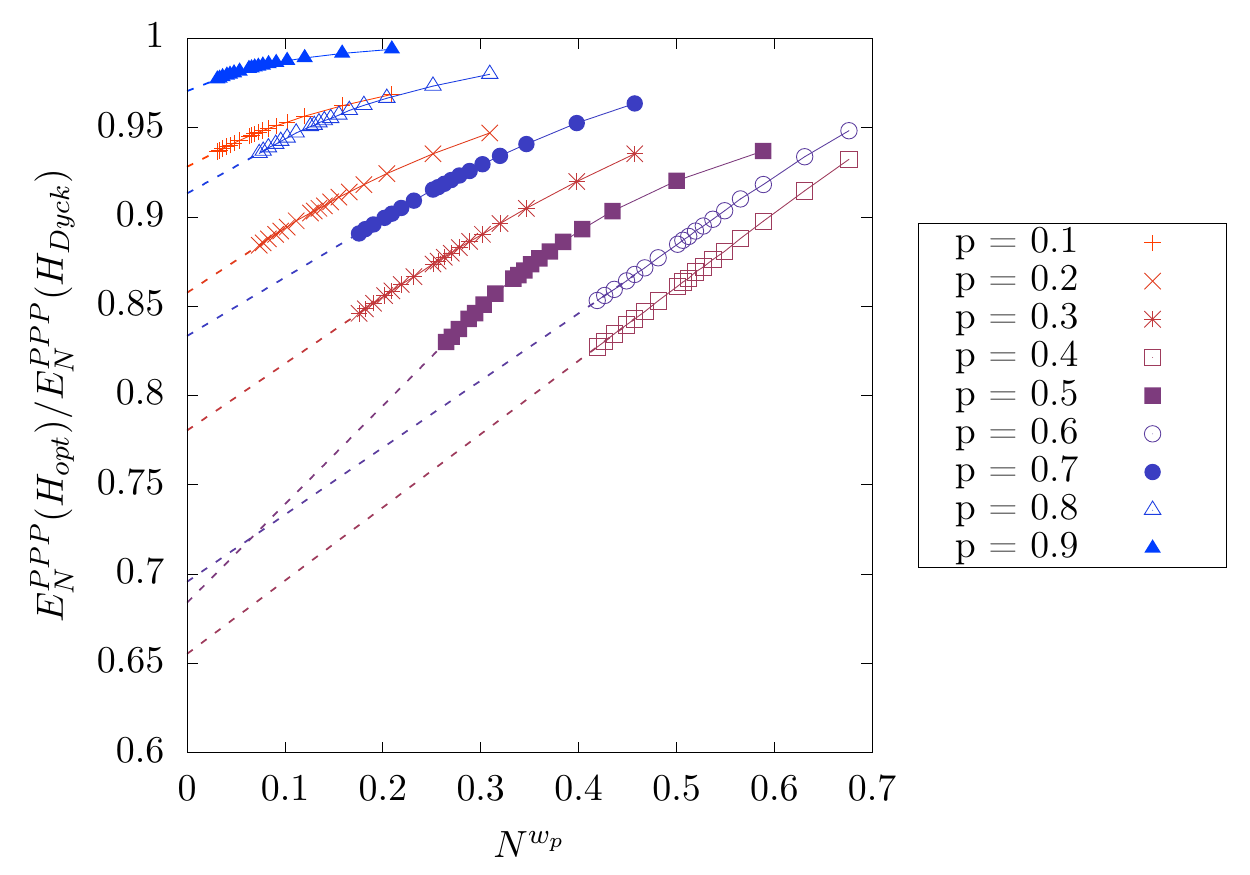}
  \end{center}
  \caption{
      Ratio of the average cost of optimal matchings over that of Dyck
      matchings as a function of $N^{w_{p}}$, where $w_{p} = - | p-\frac{1}{2} |$.
      For $p=\frac{1}{2}$, the ratio is plotted against $1/\log{N}$.
      The dashed lines are the linear extrapolations for $N^{w_{p}} \rightarrow 0$.
      The number of simulated instances 
at each value of $(p,n)$ is 10000, whenever $n \leq 4000$, and
5000, whenever $n=5000$ or $6000$.
      }
  \label{img:costs}
\end{figure}

Figure~\ref{img:costs} gives a comparison between the two average
costs by plotting their ratio as a function of $N$ for various values
of $p$.
The corresponding fits seem to exclude the possibility that the limit
for large $N$ of the ratio of average costs go to zero algebraically
in $N$ (and also makes it reasonable that there are no logarithmic
factors, although this is less evident), that is, these data suggest
the content of Conjecture \ref{conj.main}.

\begin{figure}
  \centering
  \begin{center}
      \includegraphics{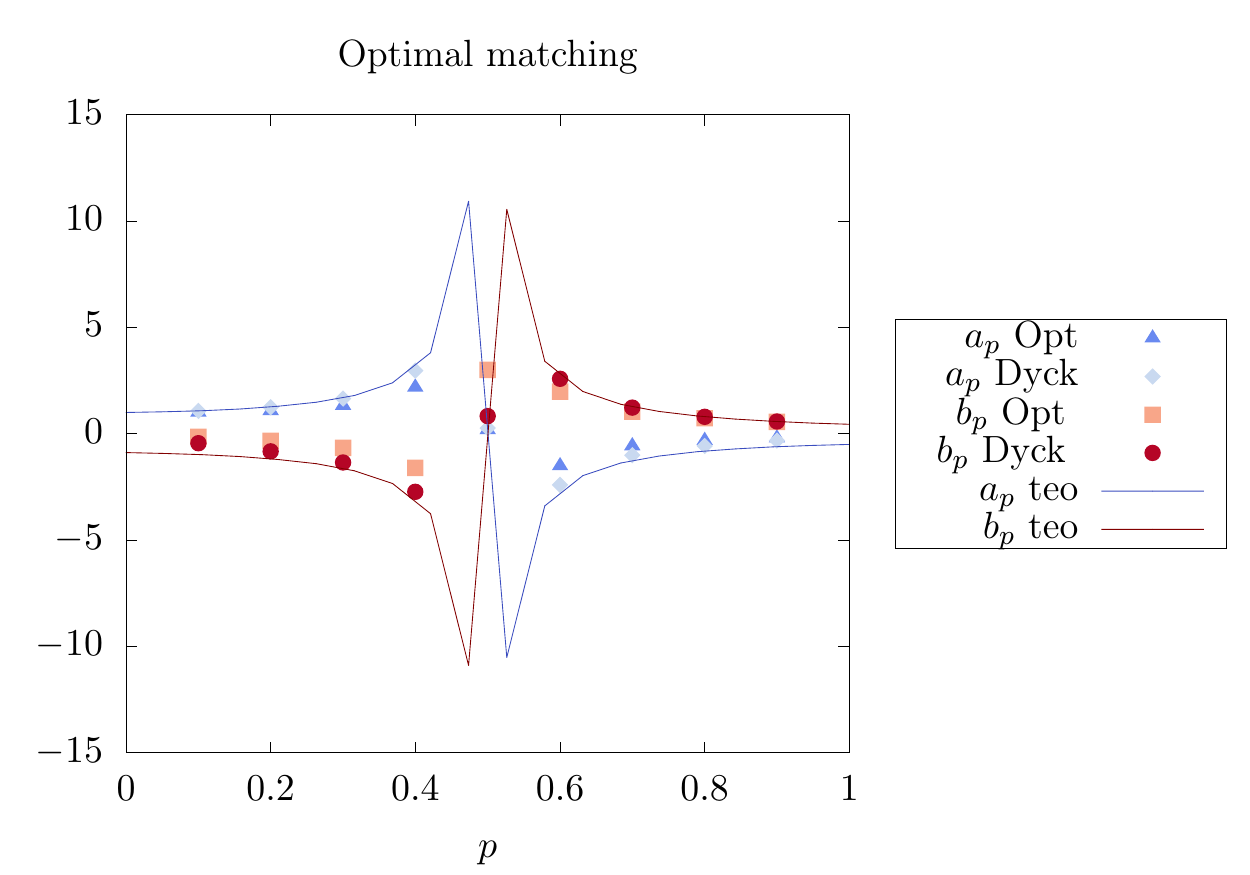}
  \end{center}
  \caption{
      Fitted scaling coefficients as a function of $p$.
  }
  \label{img:fit}
\end{figure}

In order to further test this hypothesis, we fitted the optimal
average cost to the same scaling behaviour found for the Dyck matching
average cost, i.e.
\begin{equation}
        \begin{cases}
a_p N + b_p N^{\frac{1}{2}+p} &\qquad p\neq\frac{1}{2}\\
c \, N \bigl( \log{N} + d \bigr) &\qquad p = \frac{1}{2}
        \end{cases}
\end{equation}
fixing the scaling exponents and aiming to compute the scaling coefficients.
Notice that the term $a_p N$ is leading for $0<p<\frac{1}{2}$, while $b_p N^{\frac{1}{2}+p}$ is leading for $\frac{1}{2}<p<1$.
Figure~\ref{img:fit} summarizes the fitted parameters.

For the Dyck matching, the fitted parameter for the leading scaling
coefficient agrees perfectly with the computed coefficient, as
expected.  The coefficient of the subleading term seems to agree with
the computed coefficient in a less precise way, probably due to
stronger higher-order corrections.  For the optimal matching, the
fitted coefficients behave qualitatively as the coefficients that we
have computed for the Dyck case, but the agreement is visibly not
quantitative.  The fit seems to confirm the hypothesis that the two
average costs have the same scaling exponents with different
coefficients.

To completely confirm such hypothesis, we suggest that lower bounds
for the cost could be computed.  We expect such lower bounds to share
the same scaling as that found in this paper, but with different
constants.  We leave such matter open for future work.

\section{Conclusion}

In this paper we have addressed the random Euclidean assignment
problem in dimension 1, for points chosen in an interval, with a 
cost function which is an increasing and concave power law, that is
$c(|x|)=|x|^p$ for $0<p<1$. 
We introduced a new special matching configuration associated to an
instance of the problem, that we called the \emph{Dyck matching}, as
it is produced from the Dyck bridge that describes the interlacing of
red and blue points on the domain.

\begin{figure}
  \centering
  \includegraphics{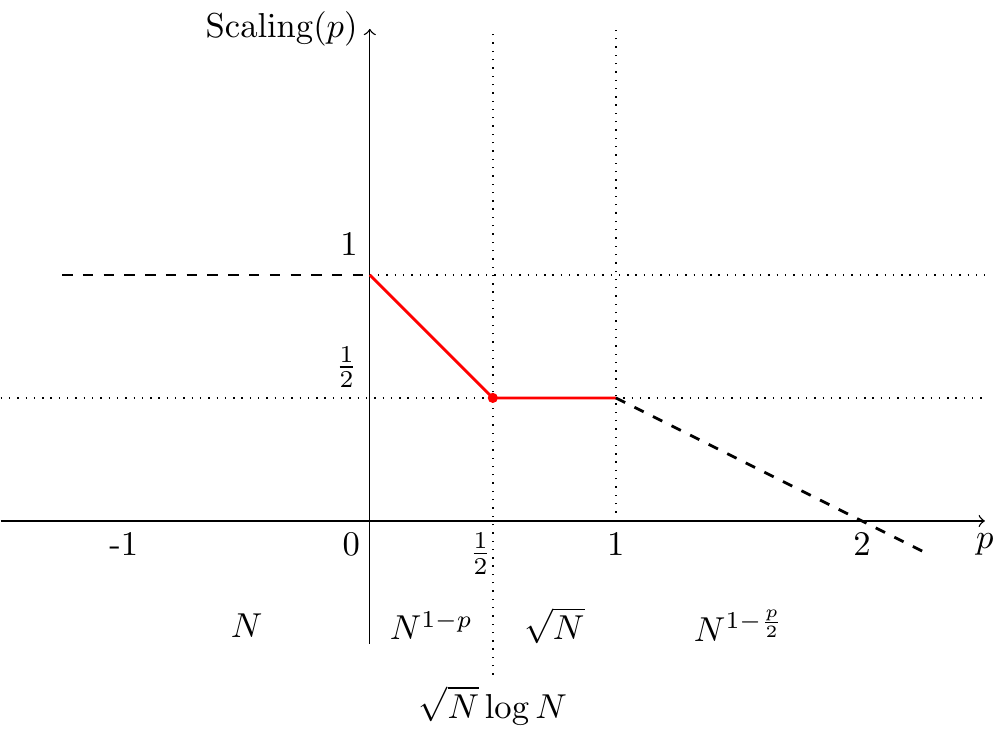}
  \caption{Scaling exponent of the average optimal cost as a function
    of $p$, in the case of cost function $c(|x|)=+|x|^p$ (that is,
    attractive case for $p>0$ and repulsive case for $p<0$).
    In red (solid line), our conjectured result.
    In black (dashed line), results from \cite{Caracciolo:177}.
  Notice that we rescaled our result by a factor $(2N+1)^{-p}$,
  i.e.\ we plotted the result for $\Omega=[0,1]$.}
  \label{img:allP}
\end{figure}

As this is a deterministic configuration, described directly in terms
of the instance, instead of involving a complex optimisation problem, 
this configuration is much more tractable than the optimal
matching. On top of this fact, we can exploit a large number of nice
facts, from combinatorial enumeration, which provides us also with
several results which are exact at finite size, this being, to some
extent, surprising. In particular, we computed the average cost of
Dyck matchings under a particular choice of probability measure (the
one in which the spacings among consecutive points are
i.i.d.\ exponential variables).
Finally, we performed numerical simulations that suggest that the
average cost of Dyck matchings has the same scaling behaviour of the
average cost of optimal matchings. We leave this claim as a conjecture.
A promising way to prove this conjecture seems to be that of providing
a lower bound on the average cost of optimal matchings with the same
scaling as our upper bound, by producing ``sufficiently many'' or
``sufficiently large'' sets of edge-lengths which must be taken by the optimal solution.
If we assume our conjecture, this result allows to fill in a missing
portion in the phase diagram of the model, for what concerns the
scaling of the average optimal cost as a function of $p$ (see
Figure~\ref{img:allP}). These new facts highlight a much richer
structure that what could have been predicted in light of the previous
results alone, with the concatenation of four distinct regions, and a
special point with logarithmic corrections.

The case of uniformly spaced points needs further analysis in the
region $p<1/2$. One can define an interpolating family of independent
spacing models, which encompasses both the ES and US cases, by taking
as function $f(s)$ the Gamma distribution $\alpha g_\alpha(\alpha s)$,
for $\alpha>0$. For example, when $\alpha$ is an integer, each $s_{i}$
is distributed as a sum of $\alpha$ i.i.d. exponential random
variables, each with mean $\alpha^{-1}$. The ES case is, of course,
$\alpha=1$, while, due to the central limit theorem, the US model is
the limit as $\alpha$ tends to infinity.  This generalised model
appears to be treatable with the same technique that we employed for
the pure ES case whenever $\alpha$ is a half-integer: the generating
function of the average cost can be computed exactly in this case, and
involves more and more complicated hypergeometric functions as
$\alpha$ grows (namely, if $\alpha=k/2$, we have a ${}_kF_{k-1}$
hypergeometric function).  Performing singularity analysis
over such functions is a challenging task, which builds on classical
results on generalised hypergeometric functions (due to 
N{\o}rlund and B\"uhring), that we leave for future
work.

On a parallel but distinct research direction, Dyck matchings may
provide a fruitful framework to study other interesting features of
the assignment problem, such as the ground-state degeneracy at $p=1$
or the properties of the cycle decomposition of optimal matchings,
w.r.t.\ the natural ordering along the domain segment.

\appendix

\section{The coefficients $v_{N,k}$} 
\label{ap.vnk}

The goal of this section is to compute the coefficients $v_{N,k}$,
crucially used in the calculation of the average cost of the Dyck
matching, starting with equation (\ref{eq.sumCost}). These
coefficients count, among the $N$ edges of all the possible
$\binom{2N}{N}$ Dyck matchings on $2N$ points, the number of edges $e$
of length $\|e\|=2k+1$. That is, $v_{N,k} \frac{(N-1)!^2}{2 (2N-1)!}$ is the
probability that, taking a random Dyck matching $\pid$ uniformly, and
an edge $e \in \pid$ uniformly, we have $\|e\|=2k+1$.

Dyck matchings correspond to Dyck `bridges', w.r.t.\ the notation
introduced in Section \ref{ssec.basicprops}.  We proceed with the
calculation by first computing the analogous quantity on a restricted
ensemble, associated to Dyck `excursions' (that is, the ordinary Dyck
paths), which are Dyck bridges satisfying the extra condition
$\sum_{i=1}^{j} \sigma_{i} \geq 0$ for all $1 \leq j \leq 2N$.

In the whole class of Dyck paths of length $2N$ there are
\begin{equation} \label{eq:rnk}
  r_{N, k} :=  \frac{N-k+1}{2}\, C_k \, C_{N-k} = \frac{1}{2}\, C_k\, B_{N-k}
\end{equation}
edges of length $\|e\|=2k+1$ (\url{http://oeis.org/A141811}), with 
$0\leq k \leq n-1$.
These numbers obey the recursion relation, which determines them
univocally (together with the initial conditions)
\begin{equation} \label{eq:rnk_rec} 
    r_{N, k}= C_k C_{N-k-1} +
  \sum_{m=0}^{N-1} \left[ r_{m, k}\, C_{N-m-1} + r_{N-m-1, k}\, C_m\right]
  \, .
\end{equation}
The recursion can be understood in terms of a first-return decomposition.
If we decompose the path into its \emph{first return}, i.e.\ the portion between its
left endpoint and its first zero (say at position $2m+2$, $0<m<n-1$),
and into its \emph{tail}, i.e.\ the remaining portion of the path on
the right, then:
\begin{itemize}
\item the first term counts all the paths in which the link between
  the first step and the first zero is of the required length. The
  multiplicity of paths in which this situation arise is given by all
  the possible paths composing the first excursion times all the
  possible paths composing the tail;
\item the sum counts, for all the possible positions of the first
  zero, the possible links of the required length hidden in the first
  excursion or in the tail of the path.
    To count links of the required length hidden in the first
    excursion, one can use $r_{m,l}$ itself, times all the possible
    tails $C_{n-m-1}$. The tail case is symmetric.
\end{itemize}

It is easy to prove by induction that
\begin{equation}
  r_{N, k} = C_k\, R_{N-k-1}
\end{equation}
and the recursion reduces to
\begin{equation}
  R_s = C_{s} + 2 \sum_{m=1}^{s-1} C_{s-m} \,R_{m-1}\, .
\end{equation}
By introducing the generating function
\begin{equation}
  R(z) := \sum_{n\ge 0} R_n\, z^n
\end{equation}
we get the equation
\begin{equation}
  R(z) = C(z) + 2\, z\, C(z)\, R(z)
\end{equation}
and therefore
\begin{equation} \label{eq:R(z)} 
    \begin{split}
        R(z) &= \frac{C(z)}{1 - 2\, z\, C(z)} = C(z) B(z) = \frac{1}{2\, z} \left[ \frac{1}{\sqrt{1- 4\, z}} -1 \right] \\
 &= - \frac{1}{2\, z} + \frac{1}{2\, z} + \sum_{n\ge 0} \frac{B_{n+1}}{2}\, z^n = \sum_{n\ge 0} \frac{B_{n+1}}{2}\, z^n
    \end{split} 
\end{equation}
indeed
\begin{equation}
\sum_{k=1}^n C_k B_{n-k} = \frac{B_{n+1}}{2}\, .
\end{equation}
It follows that
\begin{equation}
  R_{N-k-1} = \frac{B_{N-k}}{2}
\end{equation}
as announced.

The preliminary computation of the $r_{N,k}$ coefficients suggests to
use the same technique for the $v_{N,k}$, and provides an ingredient
to write a recursion for the $v_{N,k}$:
\begin{equation}
  \begin{split}
    v_{N,k} = & \, 2 \, C_k \, B_{N-k-1} + 2\, \sum_{m=0}^{N-1} \left( r_{m,k}\, B_{N-m-1} + v_{N-m-1,k} \, C_m\right) \\
    = & \, 2 \, C_k \, B_{N-k-1} + \sum_{m=k+1}^{N-1} C_k \, B_{m-k} \, B_{N-m-1} + 2 \, \sum_{m=k+1}^{N-1}  v_{m,k} \, C_{N-m-1} \\
    = & \, 2 \, C_k \, B_{N-k-1} + \sum_{m=1}^{N-k-1} C_k \, B_{m} \, B_{N-k-m-1} + 2 \, \sum_{m=1}^{N-k-1}  v_{m+k,k} \, C_{N-k-m-1} 
  \end{split}
\end{equation}
and if we again set
\begin{equation}
  v_{N,k} :=  C_k V_{N-k-1}
\end{equation}
we get
\begin{equation}
  \begin{split}
    V_s = & \, 2\, B_s + \sum_{m=1}^s B_m \, B_{s -m} + 2\, \sum_{m=1}^s 
      V_{m-1}\, C_{s -m}\\
    = & \, B_s + \sum_{m=0}^s B_m \, B_{s -m} + 2\, \sum_{m=1}^s V_{m-1}\, C_{s -m} \\
    = & \, B_s + 4^s + 2\, \sum_{m=1}^s V_{m-1}\, C_{s -m} \, .
  \end{split}
\end{equation}
We introduce now the generating function
\begin{equation}
  V(z) := \sum_{k\ge 0} V_k \, z^k
\end{equation}
to get the relation
\begin{equation}
  V(z) =  \, \frac{1}{\sqrt{1- 4\,z}} + \frac{1}{1- 4\, z} + 2\, z \, \frac{1-\sqrt{1- 4\,z} }{2\,z}\, V(z)
\end{equation}
so that
\begin{equation}
  V(z) = \frac{1}{1- 4\, z} + \frac{1}{(1- 4\, z)^\frac{3}{2}}\, =  \sum_{k\ge 0} \left[4^k + \frac{(2 k+1)!}{(k!)^2} \right] \, z^k \, ,
\end{equation}
which is our seeked result. We can finally check that the recursion
above is indeed satisfied, as
\begin{equation}
  \begin{split}
    v_{N,k} = & \, C_k \left[4^{N-k-1} + \frac{(2 N- 2k -1)!}{[(N-k-1)!]^2} \right] \\
    =  & \, C_k \left[4^{N-k-1} + \frac{(N-k)^2}{2\,(N-k)}  \, B_{N-k} \right] \\
    = & \, 4^{N-k-1}\, C_k + (N-k) \, r_{N,k} \, .
  \end{split}
\end{equation}

\section{Singularity analysis} \label{ap:asy}

Singularity analysis is a technique that allows to extract information on
the coefficients of a generating function $f(z)$ when an explicit
series expansion
around $z=0$ is not available. Roughly speaking, two main principles hold
(see e.g.\ \cite[pg.\,227]{flajolet2009analytic}):
\begin{enumerate}
\item the moduli of the singularities of $f$ dictate the asymptotic
  exponential growth of its coefficients. If $z=a$ is a singularity of
  $f(z)=\sum_{n\geq0}f_nz^n$, then $f_n \sim |a|^{-n}$;
\item the nature of the singularities of $f$ dictate the asymptotic
  sub-exponential growth of its coefficients, i.e. they determine the
  (typically polynomial or logarithmic) function $\theta(n)$ such that
  $f_n \sim |a|^{-n} \, \theta(n)$.
\end{enumerate}
We will specialise this analysis to the case of a single singularity,
along the real positive axis, which is pertinent to series with
positive coefficients, and no oscillatory behaviour.  Generalizations
of these principles (and of the related theorem below) for the case of
multiple singularities at the same radius hold as well, but in our
case are not relevant and will not be discussed.

The main result that we are going to need is a theorem (see
\cite[thm.\ VI.4]{flajolet2009analytic}) that states that if $f(z)$ is
a ``well-behaved'' complex function analytic in 0, with a singularity
at $z=\zeta + i0$ such that
\begin{equation}
  f(z) = \sigma(z/\zeta) + o(\tau(z/\zeta))
\end{equation}
for some functions $\sigma = \sum_{n\geq0}\sigma_nz^n$ and
$\tau=\sum_{n\geq0}\tau_nz^n$ in the span of the reference set
\begin{equation}
  \mathcal{S} = \Bigl\{ (1-z)^\alpha \left( \frac{1}{z}
  \log{\frac{1}{1-z}} \right)^\beta \Bigm| \alpha,\beta \in \mathbb{C}
  \Bigr\} \, ,
\end{equation}
then
\begin{equation}
  f_n = \zeta^{-n} \sigma_n + o(\zeta^{-n} \tau_n) \, .
\end{equation}
Here ``well behaved'' means that there exists an indented disk of
radius bigger than $\zeta$, with the indentation that specifically
excludes $z=\zeta+i0$, where $f(z)$ can be analytically continued.
This means that the theorem is applicable to functions with very
general singularities (isolated poles, branch cuts, \dots), and in
particular to hypergeometric functions.


The reference set $\mathcal{S}$ is composed of functions whose
expansion can be computed exactly thanks to generalizations of the
binomial theorem.  These functions are ubiquitous in series expansions
around poles of complex functions, so that the theorem is extremely
versatile.

In our specific case of the ES model, the generating function to study
is of the form
\begin{equation}
  f(z) = s(4z) \, F\left(\begin{array}{c} a , b \\ c \end{array} \bigg\rvert \, 4z \right)
\end{equation}
where $s(z) \in \text{span}\,\mathcal{S}$ is singular at $z=1$, and
$F={}_2F_1$ is the $(2,1)$-hypergeometric function defined by
\begin{equation} \label{eq:def_hyper}
  F\left(\begin{array}{c} a , b \\ c \end{array} \bigg\rvert \, z
  \right) = \sum_{n\geq0} \frac{\Gamma(n+a)}{\Gamma(a)}
  \frac{\Gamma(n+b)}{\Gamma(b)} \frac{\Gamma(c)}{\Gamma(n+c)}
  \frac{z^n}{n!}
\ef,
\end{equation}
or, equivalently, by
\begin{align} 
\label{eq:def_hyper2}
  F\left(\begin{array}{c} a , b \\ c \end{array} \bigg\rvert \, z
  \right) &= \sum_{n\geq0} s_n
  \frac{z^n}{n!}
\ef;
\\
\frac{s_{n+1}}{s_n} &= \frac{(a+n)(b+n)}{c+n}
\ef;
\qquad
s_1 = 1
\ef.
\end{align}
To expand and study the hypergeometric function around $z=1$, a celebrated
`inversion formula' due to Gauss is available
\begin{equation}
  \begin{split}
    F\left(\begin{array}{c} a , b \\ c \end{array} \bigg\rvert \, z \right) = &\frac{\Gamma(c)\Gamma(c-a-b)}{\Gamma(c-a)\Gamma(c-b)}  F\left(\begin{array}{c} a , b \\ a+b+1-c \end{array} \bigg\rvert \, 1-z \right) +\\
    & \frac{\Gamma(c)\Gamma(a+b-c)}{\Gamma(a)\Gamma(b)} (1-z)^{c-a-b}
    F\left(\begin{array}{c} c-a , c-b \\ c+1-a-b \end{array} \bigg\rvert \,
      1-z \right)
  \end{split}
\end{equation}
This formula restates the seeked expansion around $z=0$ in terms of an
expansion near the singularity at $z=1$.
As the hypergeometric function is analytic in $z=0$, the singular
behaviour at $z=1$ of the right-hand side combination is described 
by the power-law prefactors in the inversion formula.

In our specific case, $a=\frac{p+1}{2}$, $b=\frac{p+2}{2}$ and $c=2$
with $p \in [0,1]$, giving 
$c-a-b = \frac{1-2p}{2} \in [-\frac{1}{2}, \frac{1}{2}]$.
Thus, the leading terms of the expansion of $f(z)$ are:
\begin{multline}
  f(z) = s(4z) \bigg[
    \frac{\Gamma(c)\Gamma(c-a-b)}{\Gamma(c-a)\Gamma(c-b)} +
    \frac{\Gamma(c)\Gamma(a+b-c)}{\Gamma(a)\Gamma(b)} (1-4z)^{c-a-b}
    \bigg. \\
  \bigg. + \Theta((1-4z), (1-4z)^{c-a-b+1}) \bigg].
\end{multline}
The above expression is valid only for $p\neq\frac{1}{2}$.
A limit procedure combines the diverging $\Gamma$'s and the $(1-4z)$ term to give
\begin{multline}
  f(z) = s(4z) \bigg[ - \frac{1}{\Gamma(\frac{3}{4})
      \Gamma(\frac{5}{4})} \left( \ln(1-4z) +2\gamma
    +\psi_0\left(\smfrac{3}{4}\right) + \psi_0\left(\smfrac{5}{4}\right)
    \right) 
    \bigg. \\
  \bigg. + \Theta((1-4z) \ln(1-4z)) \bigg].
\end{multline}
where $\gamma$ is the Euler-Mascheroni constant and $\psi_0$ is the
digamma function.  The limit is to be performed with care: each term
must be written as a function of $\epsilon=p-\frac{1}{2}$ and expanded
in powers series.  The expansion of the hypergeometric functions must
be performed using their definition.  When everything is expanded,
$o(\epsilon)$ are discarded taking the limit $\epsilon\rightarrow0$,
and the leading terms in the $(1-4z)$ are found by taking $n=0$ in the
sum of the hypergeometric function definition.

\bibliographystyle{unsrt}

\begin{thebibliography}{10}

\bibitem{Orland1985}
Henri Orland.
\newblock {Mean-field theory for optimization problems}.
\newblock {\em Le Journal de Physique - Lettres}, 46(17):773--770, 1985.

\bibitem{Mezard1985}
Marc M{\'{e}}zard and Giorgio Parisi.
\newblock {Replicas and optimization}.
\newblock {\em Journal de Physique - Lettres}, 46(17):771--778, 1985.

\bibitem{Mezard1986a}
Marc M{\'{e}}zard and Giorgio Parisi.
\newblock {Mean-field equations for the matching and the travelling salesman
  problems}.
\newblock {\em Europhysics Letters}, 2(12):913--918, 1986.

\bibitem{Caracciolo:168}
Sergio Caracciolo, Matteo~P. D'Achille, Enrico~M. Malatesta, and Gabriele
  Sicuro.
\newblock Finite size corrections in the random assignment problem.
\newblock {\em Phys. Rev. E}, 95:052129, 2017.

\bibitem{lovasz2009matching}
L\'{a}szl\'{o} Lov\'{a}sz and Michael~D. Plummer.
\newblock {\em Matching Theory}, volume 367 of {\em AMS Chelsea Publishing
  Series}.
\newblock North-Holland; Elsevier Science Publishers B.V., 2009.

\bibitem{Parisi-conj}
Giorgio Parisi.
\newblock A conjecture on random bipartite matching.
\newblock {\em {\tt arXiv:cond-mat/9801176}}, 1998.

\bibitem{CopSor}
D.~Coppersmith and G.~Sorkin.
\newblock Constructive bounds and exact expectations for the random assignment
  problem.
\newblock {\em Random Structures and Algorithms}, 15:113---144, 1999.

\bibitem{Aldous2001}
David~J. Aldous.
\newblock {The $\zeta$(2) limit in the random assignment problem}.
\newblock {\em Random Struct. Algorithms}, 2:381--418, 2001.

\bibitem{Nair2005}
Chandra Nair, Balaji Prabhakar, and Mayank Sharma.
\newblock {Proofs of the Parisi and Coppersmith-Sorkin random assignment
  conjectures}.
\newblock {\em Random Struct. Algorithms}, 27(4):413--444, 2005.

\bibitem{Linusson}
Svante Linusson and Johan Wastlund.
\newblock {Completing a $(k-1)$-assignment}.
\newblock {\em {Combinatorics Probability \& Ccomputing}}, 16(4):621--629,
  {JUL} 2007.

\bibitem{Mezard1988}
Marc M{\'{e}}zard and Giorgio Parisi.
\newblock {The euclidean matching problem}.
\newblock {\em Journal de Physique}, 49:2019--2025, 1988.

\bibitem{Ajtai}
Mikl{\'{o}}s Ajtai, J{\'{a}}nos Koml{\'{o}}s, and Gabor Tusn{\'{a}}dy.
\newblock {On optimal Matchings}.
\newblock {\em Combinatorica}, 4(4):259--264, 1984.

\bibitem{Caracciolo:158}
Sergio Caracciolo, Carlo Lucibello, Giorgio Parisi, and Gabriele Sicuro.
\newblock Scaling hypothesis for the euclidean bipartite matching problem.
\newblock {\em Phys. Rev. E}, 90:012118, 2014.

\bibitem{Caracciolo:162}
Sergio Caracciolo and Gabriele Sicuro.
\newblock {Scaling hypothesis for the Euclidean bipartite matching problem. II.
  Correlation functions}.
\newblock {\em Phys. Rev. E}, 91:062125, 2015.

\bibitem{Caracciolo:163}
Sergio Caracciolo and Gabriele Sicuro.
\newblock Quadratic stochastic euclidean bipartite matching problem.
\newblock {\em Phys. Rev. Lett.}, 115(23):230601, 2015.

\bibitem{Ambrosio2016}
Luigi Ambrosio, Federico Stra, and Dario Trevisan.
\newblock {A PDE approach to a $2$-dimensional matching problem}.
\newblock {\em Probab. Theory Relat. Fields}, pages 1--45, 2018.

\bibitem{Bobkov2016}
Sergey Bobkov and Michel Ledoux.
\newblock {One-dimensional empirical measures, order statistics, and
  Kantorovich transport distances}.
\newblock {\em Memoirs of the AMS}, 261(1259), 2019.

\bibitem{Caracciolo:172}
Sergio Caracciolo, Matteo~P. D'Achille, and Gabriele Sicuro.
\newblock Anomalous scaling of the optimal cost in the one-dimensional random
  assignment problem.
\newblock {\em J. Statist. Phys.}, 174(4):846--864, 2019.

\bibitem{Talagrand2018}
Michel Talagrand.
\newblock Scaling and non-standard matching theorems.
\newblock {\em C. R. Acad. Sci. Paris, Ser. I}, 356(6):692--695, 2018.

\bibitem{Caracciolo:160}
Sergio Caracciolo and Gabriele Sicuro.
\newblock On the one dimensional euclidean matching problem: exact solutions,
  correlation functions and universality.
\newblock {\em Phys. Rev. E}, 90:042112, 2014.

\bibitem{Caracciolo:169}
Sergio Caracciolo, Matteo~P. D'Achille, and Gabriele Sicuro.
\newblock Random euclidean matching problems in one dimension.
\newblock {\em Phys. Rev. E}, 96:042102, 2017.

\bibitem{Caracciolo:177}
Sergio Caracciolo, Andrea Di~Gioacchino, Enrico~M. Malatesta, and Luca~G.
  Molinari.
\newblock Selberg integrals in 1d random euclidean optimization problems.
\newblock {\em J. Stat. Mech.}, page 063401, 2019.

\bibitem{McCannRobert1999}
{McCann Robert}.
\newblock {Exact solutions to the transportation problem on the line}.
\newblock {\em Proc. R. Soc. A: Math., Phys. Eng Sci}, 455:1341--1380, 1999.

\bibitem{Caracciolo:159}
Elena Boniolo, Sergio Caracciolo, and Andrea Sportiello.
\newblock Correlation function for the grid-poisson euclidean matching on a
  line and on a circle.
\newblock {\em J. Stat. Mech.}, 11:P11023, 2014.

\bibitem{Higgs2000}
Paul~G. Higgs.
\newblock {RNA secondary structure: physical and computational aspects}.
\newblock {\em Q. Rev. Biophys.}, 33(03):199--253, 2000.

\bibitem{Orland2002}
Henri Orland and Anthony Zee.
\newblock {RNA folding and large N matrix theory}.
\newblock {\em Nuclear Physics B}, 620(3):456--476, 2002.

\bibitem{Vernizzi}
Graziano Vernizzi, Henri Orland, and Anthony Zee.
\newblock {Enumeration of RNA Structures by Matrix Models}.
\newblock {\em Phys. Rev. Lett.}, 94:168103, 2005.

\bibitem{Nechaev2013}
S.~K. Nechaev, A.~N. Sobolevski, and O.~V. Valba.
\newblock {Planar diagrams from optimization for concave potentials}.
\newblock {\em Phys. Rev. E}, 87(1):012102, 2013.

\bibitem{Bundschuh2002}
R~Bundschuh and Terence Hwa.
\newblock {Statistical mechanics of secondary structures formed by random RNA
  sequences}.
\newblock {\em Phys. Rev. E}, 65(3):031903, 2002.

\bibitem{Muller2003rna}
M.~M{\"u}ller.
\newblock {Statistical physics of RNA folding}.
\newblock {\em Phys. Rev. E}, 67:021914, 2003.

\bibitem{Higgs1996}
Paul~G. Higgs.
\newblock {Overlaps between RNA secondary structures}.
\newblock {\em Phys. Rev. Lett.}, 76(4):704, 1996.

\bibitem{Pagnani2000}
A~Pagnani, G~Parisi, and F.~Ricci-Tersenghi.
\newblock {Glassy Transition in a Disordered Model for the RNA Secondary
  Structure}.
\newblock {\em Phys. Rev. Lett.}, 84(9):2026--2029, 2000.

\bibitem{Marinari2002}
Enzo Marinari, Andrea Pagnani, and Federico Ricci-Tersenghi.
\newblock {Zero-temperature properties of RNA secondary structures}.
\newblock {\em Phys. Rev. E}, 65:041919, 2002.

\bibitem{PitYorguide}
Jim Pitman and Marc Yor.
\newblock {A guide to Brownian motion and related stochastic processes}.
\newblock {\em {\tt arXiv:1802.09679}}, 2018.

\bibitem{delonLocalMatchingIndicators2012}
J.~Delon, J.~Salomon, and A.~Sobolevski.
\newblock Local {{Matching Indicators}} for {{Transport Problems}} with
  {{Concave Costs}}.
\newblock {\em SIAM Journal on Discrete Mathematics}, 26(2):801--827, 2012.

\bibitem{ConcreteMath}
R.L. Graham, D.E. Knuth, and O.~Patashnik.
\newblock {\em Concrete mathematics}.
\newblock Addison-Wesley, Reading, MA, 1989.

\bibitem{ArtinBook}
E.~Artin.
\newblock {\em The Gamma Function}.
\newblock Holt, Rinehart, Winston, Inc., 1964.

\bibitem{flajolet2009analytic}
Philippe Flajolet and Robert Sedgewick.
\newblock {\em Analytic combinatorics}.
\newblock Cambridge University Press, 2009.

\end{thebibliography}

\end{document}